\documentclass[reqno]{amsart}

\usepackage{amssymb,epsfig,graphics}
\usepackage{amsmath}
\usepackage{amscd}
\usepackage{verbatim}
\input xypic

\allowdisplaybreaks

\newtheorem{thm}{Theorem}[section]
\newtheorem{prop}[thm]{Proposition}

\theoremstyle{definition}
\newtheorem{definition}{Definition}
\newtheorem{example}{Example}

\def\C{\mathbb C}
\def\R{\mathbb R}
\def\Z{\mathbb Z}

\def\Q{\mathbb Q}
\def\T{\mathbb T}

\def\r{\rangle}
\def\l{\langle}

\def\wt{\widetilde}

\def\tt{\tilde t}
\def\tT{\tilde T}
\def\ts{\tilde s}

\def\tomaps{\leftarrow\!\shortmid}

\begin{document}

\title[Almost periodic expansions]
{Computing with almost periodic functions}

\author{R.V. Moody}

\author{M. Nesterenko}

\author{J. Patera}


\begin{abstract}
The paper develops a method for discrete computational Fourier analysis of functions defined on
quasicrystals and other almost periodic sets. A key point is to build the analysis around the emerging
theory of quasicrystals and diffraction in the setting on local hulls and dynamical
systems.  Numerically computed approximations arising in this way are
built out of the Fourier module of the quasicrystal in question,
and approximate their target functions uniformly on the entire infinite space.

The methods are entirely group theoretical, being based on finite groups
and their duals, and they are practical and computable.
Examples of functions based on the standard
Fibonacci quasicrystal serve to illustrate the method (which is applicable
to all quasicrystals modeled on the cut and project formalism).
\end{abstract}

\maketitle
\section{Introduction}\label{Introduction}
In this paper we consider the problem of discrete methods for dealing with functions
that are intrinsically almost periodic, but not actually periodic. Quasicrystals, quasicrystalline
photonic crystals, Faraday wave experiments, and other physical phenomena
arising from the interaction of incommensurate frequencies, all display
the features of almost periodicity. As a typical example one may think of a potential field of a physical quasicrystal.
The salient features of quasicrystals are highly structured long-range order
(represented by pure point or near pure point diffraction) but no periodic order.
Thus the potential is not a periodic function, but rather belongs to the domain of almost periodic functions.

Here we put forth a method for finite discrete analysis of almost periodic functions
that has the following main features:
\begin{itemize}
\item It is entirely based on group theoretical methods, primarily finite groups and their duals.
\item The discretely computed Fourier approximants are themselves almost periodic and uniformly
approximate their target functions over their entire domains.
\item The  Fourier frequencies involved in the approximation lie in the module
of Fourier frequencies of the target function.
\end{itemize}

A standard approach to modeling such structures is to take a finite part of it,
impose periodic boundary conditions, rationalize and reduce the object to a periodic approximant,
and then apply usual crystallography.
Although this type of periodization is used routinely and successfully for many modelling problems
in the theory of quasicrystals, it is not entirely satisfactory.
Almost periodic order goes beyond periodic order in fundamental ways,
its essence appearing as a underlying incommensurability which pervades every part of the theory.
For instance, a key feature of quasicrystals appears in the Fourier module which parameterizes the Bragg
spectrum and always has rank higher than (typically double) the dimension of the ambient space of the quasicrystal.
Periodization destroys this, and by its nature can only produce results that can fit data over the finite range
specified by the
imposed periodization boundaries, whereas the essence of the material is that its order is long-range.
The present paper does not involve any periodization and avoids
these issues.

The theory of almost periodic functions was initiated  by H.~Bohr
\cite{bohr}, on the basis of earlier work on uniform approximation
of functions by trigonometric polynomials by P. Bohl. It was greatly
extended by the work of A.~Besicovitch \cite{besic}, S.~Bochner,
J.~von Neumann, \cite{Boch,Bochn,BochNeu}, N.~Wiener \cite{Wien},
H.~Weyl \cite{Weyl}, and others
\cite{burckel,levitan-zhikov,amerio-prouse}. The advent of
quasicrystals and aperiodic tilings instigated a revival of the
field, and led to extensive study of the cut and project formalism
and the theory of pure point diffraction \cite{Meyer, RVM, Hof,
Schlott} which have become the mainstays of experimentalists and
theorists alike. An important component of this is the use of
dynamical systems and dynamical hulls. These are ideally suited to
the phenomenon of almost periodicity, which appears in the dynamics
as recurrence, and provide a natural setting for the Fourier
analysis that is used to study it.

Although our study is of almost periodic functions, their importance in the subject
of quasicrystals is that they arise from functions whose behaviour is dominated
by the local environments of the quasicrystal in question. The way in which this happens is
mostly taken for granted, but in fact there are some interesting assumptions involved,
and for this reason we begin by formally defining local functions with respect to a
given point set $\Lambda$, and showing how it is that they are connected with almost periodic functions.

We can create a {\it dynamical hull\/} $(X,\mu)$ from $\Lambda$.
This is a compact space that arises from $\Lambda$ and its translations,
and it lies at the heart of the Fourier analysis of local functions on $\Lambda$.
One assumes (in many important cases it is forced) a probability measure
$\mu$ on which $\R^d$ acts in a measure preserving way.
A local function $f$ lifts to some new function $F$ on $(X,\mu)$,
and this is an $L^2$ function. Now the analysis of $f$ can be related directly to the analysis of $F$,
and for this we have a powerful tool
in the form of the action of $\R^d$ on $L^2(X,\mu)$, which is unitary. All of this material is explained in \S2.

To go further, we next place ourselves in the situation of the cut and project formalism,
which is the standard method of modeling used in the study of quasicrystals.
The set $\Lambda$ is now assumed to be a model set (cut and project set).
In this setting the hull $(X,\mu)$, and more particularly $L^2(X,\mu)$,
can be described explicitly in terms of a higher dimensional
torus\footnote{More generally a compact Abelian group}(higher dimensional periodicity!),
and it is a straightforward matter to carry out Fourier analysis of $F$.
It is the restriction of this Fourier analysis back to $f$ that provides the required Fourier analysis of $f$.
By its very nature this is almost periodic and captures the full aperiodic nature of $\Lambda$,
including the correct Fourier module in which physical information actually appears. This is the content of \S 3.
Readers familiar with the cut and project method who do not wish to go through the theory of local hulls and
local functions may read \S\ref{cpsMSTP} and \S\ref{fouriercoefficients},
and then continue starting at \S\ref{discretization}.

This theoretical analysis is based on higher dimensional structures that are not explicitly computatable,
as well as the usual array of countably many Fourier coefficients, each of which is the outcome of integration.
To be a practical tool, the analysis has to be reduced to finitely many objects that are explicitly computable
entirely in the context of the given function $f$. The resulting approximants are trigonometric polynomials
(quasi-periodic functions) whose frequencies come from the
Fourier module of the original function.  In \S4 we outline the method of discretization,
which depends primarily on the construction of a refinement lattice of the lattice
of the cut and project scheme and, along with it, its dual lattice. Together these
produce two finite groups which are in $\Z$-duality to each other. The selection
of data points and appropriate Fourier frequencies is governed by these two groups.

This is best illustrated by examples and for this we have chosen two local functions
based on one of the famous Fibonacci point sets. This has the advantage of being straightforward
to construct and easy to visualize, while at the same time containing all the essential features
of more general model sets. \S5 prepares the mathematics of the Fibonacci cut and project scheme,
and \S6
shows some explicit computations for the particular local functions we have chosen.
As is evident from the examples, the main effort required is in the creation of the
data points. Once this is done, the same set of data points and Fourier frequencies
will work for any almost periodic function arising from the same cut and project scheme.

The results are striking in two ways. First of all, the approximating functions are remarkably good,
given the amount of data from which they are produced. Second, the approximating functions
are not just local approximations they are also \emph{global} approximations,
in the sense that they provide finite Fourier series that approximate $f$ throughout
its entire domain (namely $\R$). Of course this was to be expected, but it is impressive to see it in action.

The numerical methods we introduce here are designed to be efficient and, of course,
to utilize the inherent almost periodicity. In the development of the theory we provide
error estimates that can be worked out specifically in the cases of interest.
We note that the primary weakness in the error estimates is not one that is due to
the aperiodic nature of the problem, but one that always appears in Fourier analysis,
namely how well can one approximate a function if one uses only finitely many of its Fourier coefficients.

\bigskip

\section{Continuous functions on aperiodic point sets} \label{Cfaps}

\subsection{Local hulls}\ \label{hulls}

We work in $\R^d$, a real Euclidean space of finite dimension $d$.
For $r>0$, let $D_r(\R^d)$, be all point sets $\Lambda\subset\R^d$
for which the distance $|x-y|\geq r$ for all $x,y\in\Lambda$.
This means that $D_r$ consists of all discrete point sets with minimal separation $\geq r$ between their points.

The {\it local topology\/} on $D_r(\R^d)$ can be intuitively introduced as follows.
Two sets $\Lambda_1$ and $\Lambda_2$ of $D_r(\R^d)$ are `close' if, for some large $R$
and some small $\epsilon$, one has
\begin{equation}
\begin{aligned}\label{close}
\Lambda_1\cap B_R\ &\subset\ \Lambda_2 + B_\epsilon\,,\\
\Lambda_2\cap B_R\ &\subset\ \Lambda_1 + B_\epsilon\,,
\end{aligned}
\end{equation}
where $B_R$ (resp. $B_\epsilon$) is the ball of radius $R$ (resp. $\epsilon$)
around $0$. Thus for each point of $\Lambda_1$ within the ball $B_R$,
there is a point of $\Lambda_2$ within the distance $\epsilon$ of that point;
and vice versa. Pairs $(\Lambda_1,\Lambda_2)$ satisfying \eqref{close} are called $(R,\epsilon)$-close.

\begin{figure}[h]
\includegraphics[scale=0.6]{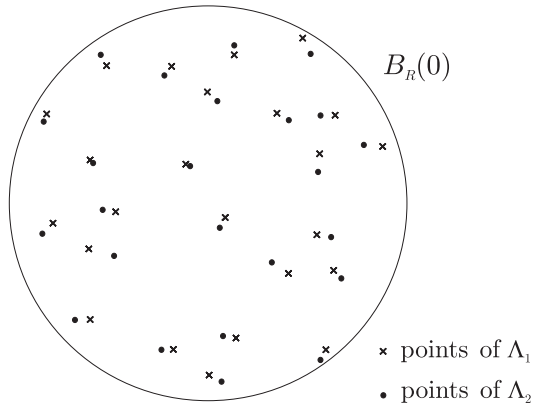}
\centering
\caption{Two sets $\Lambda_1$ and $\Lambda_2$ of $D_r(\R^d)$ that are close.}
\end{figure}

The local topology is actually a metric topology, although we make no use of this fact here.
\medskip

\noindent
\begin{definition}
For $\Lambda\in D_r(\R^d)$ the {\it local hull\/} of $\Lambda$ is
\begin{equation}\label{hull}
X(\Lambda)=\overline{\{t+\Lambda\ :\ t\in\R^d\}}\subset D_r(\R^d)\,,
\end{equation}
i.e. take all translates of $\Lambda$ and take their closure in the local topology.
\end{definition}
\smallskip

\begin{prop} \cite{RW}
The translation action of $\R^d$ on $\Lambda$ lifts to a translation action on $X(\Lambda)$.
The local hull $X(\Lambda)$ is compact and the $\R^d$-action on it is continuous. \qed
\end{prop}

\noindent
\begin{example}  Let $\Lambda$ be the lattice $\Z^d$ in $\R^d$.
Then $X(\Lambda)=\R^d/\Z^d$ is the $d$-torus (with its usual topology).
Intuitively we translate the lattice around with $\R^d$. But translation by an element of $\Z^d$
leaves $\Lambda$ invariant. So $\R^d/\Z^d$ parameterizes all distinct positions of $\Lambda$ under translation.
\end{example}

\begin{example}
Let $\Lambda$ be any Penrose tiling. Then $X(\Lambda)$ is the set of all Penrose tilings that are locally
indistinguishable from some translate of $\Lambda$. $X(\Lambda)$ contains considerably more
than just the translations of $\Lambda$. In fact  $X(\Lambda)$ consists of all Penrose tilings
based on the same pair of Penrose rhombs and the same orientations as appearing in $\Lambda$.
\end{example}

Generally one may think of $X(\Lambda)$ as some sort of local indistinguishability class of $\Lambda$.

\subsection{Continuous functions on $X(\Lambda)$}\label{CfX}

We assume that $\Lambda\subset D_r(\R^d)$ and $X(\Lambda)$ are as in \S\ref{hulls}.

Consider a function
\begin{equation}\label{function}
F : X(\Lambda)\ \longrightarrow\ \C\,.
\end{equation}
We can define from it a function
\begin{gather*}
f : \R^d\longrightarrow\ \C\,.
\end{gather*}
by
\begin{equation}\label{f(t)}
f(t)=F(t+\Lambda)\,.
\end{equation}
If $F$ is continuous then we note that for all $t_1,t_2\in\R^d$,
\begin{align*}
t_1+\Lambda,&\ t_2+\Lambda \quad \text{are close}\\
&\Longrightarrow\quad
F(t_1+\Lambda),\  F( t_2+\Lambda)\quad\text{are close}\\
&\Longrightarrow\quad
f(t_1),\  f(t_2)\quad\text{are close}\,.
\end{align*}
Thus continuity of $F$ implies  continuity of $f$, and we see that $f$ is local,
or almost periodic, with respect to $\Lambda$ in the following sense:

\begin{definition}
A function $f : \R^d\ \longrightarrow\ \C$ \
is called  \emph{local with respect to a set $\Lambda\in D_r(\R^d)$}, or
$\Lambda$-\emph{local}, if for all $\epsilon'>0$ there exist $R$ and $\epsilon$
so that whenever $t_1,t_2\in\R^d$ satisfy that $t_1+\Lambda$ and $t_2+\Lambda$ are $(R,\epsilon)$-close then
\begin{gather*}
|f(t_1)-f(t_2)|<\epsilon'\,.
\end{gather*}
\end{definition}
The intuitive meaning of this is that $f$ has the very natural property, at least from the perspective
of physical systems, that it looks very much the same at places where the local environment looks the same.
Local functions are easily seen to be continuous on $\R^d$.


Using locality,  we can go in the opposite direction. Let $\Lambda\in D_r(\R^d)$ and let $f : \R^d\ \rightarrow\ \C$
be local with respect to $\Lambda$. Define
\begin{gather*}
F: \{t+\Lambda\ :\ t\in\R^d\}\ \longrightarrow\ \C
\end{gather*}
(so that $F$ is a function on a part of $D_r(\R^d)$) by
\begin{gather*}
F(t+\Lambda)=f(t)\,.
\end{gather*}
Then $F$ is continuous on $\{t+\Lambda\ :\ t\in\R^d\}$ with respect to the local topology.
In fact it is {\it uniformly\/} continuous. The reason for this is that the continuity
condition which defines the local-ness of $f$ is based on the uniformity (i.e. the notion of $(R,\epsilon)$-closeness)
defining the local topology of $\{t+\Lambda\ :\ t\in\R^d\}$.
\smallskip

It follows that $F$ lifts uniquely to a continuous function \eqref{function} on the local hull $X(\Lambda)$.

\begin{prop}
For each local function $f$ with respect to $\Lambda$ there is a unique continuous function \eqref{function}
on the local hull, whose restriction to the orbit of $\Lambda$ is $f$. Every continuous function
on the local hull of $\Lambda$ arises in this way.  \qed
\end{prop}
Thus we see that a locality with respect to $\Lambda$ and the existence and continuity of an extension
function on $X(\Lambda)$ amount to the same thing.

\medskip

In the situation that $X(\Lambda)$ is equipped with an $\R^d$-invariant probability measure $\mu$
(i.e. a positive Borel measure $\mu$ with $\mu(t+A)=\mu(A)$ for all Borel sets and with $\mu(X(\Lambda))=1$),
the action of $\R^d$ on $X(\Lambda)$ leads to unitary action $T$ of
$\R^d$ on $L^2(X(\Lambda),\mu)$. Namely  for all $F\in L^2(X(\Lambda),\mu)$ and for all $t\in\R^d$,
$T_tF$ is the function defined by $T_tF(\Gamma)=F(-t+\Gamma)$ and with
\begin{gather*}
\l F\mid G\r :=\int_{X(\Lambda)}F\overline G\,d\mu
\end{gather*}
we have
\begin{gather*}
\l T_tF\mid T_tG\r =\l F\mid G\r\,.
\end{gather*}

In principle the spectral theory of $L^2(X(\Lambda),\mu)$ should allow one analyze
$\Lambda$-local functions $f : \R^d\rightarrow\C$ by analyzing their corresponding
functions $F$ on $L^2(X(\Lambda),\mu)$.

For one very important class of subsets $\Lambda$ this can actually be carried
out in detail -- namely the class of model sets, which we now introduce.

\section{Local functions on model sets}\label{Lfms}

\subsection{Cut and project schemes, model sets and torus parametrization}\label{cpsMSTP}

An important class of point sets $\Lambda \subset \R^d$ for which we know a considerable
amount about the corresponding hulls $X(\Lambda)$ is the class of cut and project sets,
or the model sets as they are often called \cite{Meyer, RVM}.

Consider the cut and project scheme
\begin{alignat}{3}\label{cps}
\R^d\ &\overset{||}{\longleftarrow}\quad
      &\R^d
      &\times\R^d
      &
      &\overset{\perp}{\longrightarrow}\quad\R^d\notag
      \\
      &&&\cup
      &&
      \\
      &\overset{1 - 1}{\longleftarrow}
      &&\widetilde{L}
      &&\overset{\text{dense image}}{\longrightarrow}\quad
      \notag
\end{alignat}
and a window $\Omega$. Here $\widetilde L$ is a lattice in $\R^d\times\R^d$
which is oriented so that the projections into $\R^d$ are $1-1$ and dense respectively.

In (\ref{cps}) the left-hand $\R^d$ is {\it physical space\/},
the space in which $\Lambda$ is going to lie. The right-hand $\R^d$ is {\it internal space},
the one that will be used to control the projection of the lattice $\widetilde L$ into physical space.
The image of $\widetilde L$ under projection into physical space is denoted by $L$.
Since this projection is one-one, $L \simeq \wt{L}$ as groups, so $L$ is a free Abelian group of rank $2d$,
i.e. it has a $\Z$-basis of $2d$ elements. However, it necessarily has accumulation points,
and the typical situation is that $L$ is dense in physical space.

It is convenient to use notation like $\tilde x, x, x'$ for the elements of $\widetilde{L}$
and their respective left and right projections. Then
$\tilde x = (x,x')$ where $x$ runs through $L$. This implies existence of the mapping
$(\,\cdot\,)'\ :\ L\rightarrow L' $  defined by $x\mapsto x'$, which passes from physical to internal space.

Note that this mapping $(\,\cdot\,)'$, as given, is only defined on $L$.
It cannot be extended in any canonical way to a mapping $\R^d\rightarrow\R^d$.
However it does extend canonically to the rational spans of the objects in question, and
we shall make use of this later.

We choose a subset $\Omega$ in internal space. This {\it window\/} is assumed to be compact,
equal to the closure of its interior, and to have boundary of measure 0. Using it we
define
\begin{equation}
\Lambda = \Lambda(\Omega)=\{x : \tilde{x}\in \widetilde{L}, \, x' \in \Omega  \}\, .
\end{equation}

Sets of the form $t+\Lambda(\Omega)$, $t\in\R^d$, are called {\it cut and project sets\/}
or {\it model sets\/}~\footnote{Model sets can be taken more generally with any locally
compact Abelian group as the internal space.}. In particular, for each $(x,y)\in\R^d\times\R^d$ we may define
\begin{equation}
\Lambda_{(x,y)}=x+\Lambda(-y+\Omega)\,.
\end{equation}

If $(x,y)\equiv(x',y') \mod \widetilde L$, then
$x+\Lambda(-y+\Omega)=x'+\Lambda(-y'+\Omega)$,
as can be verified directly from the definitions.
Thus these model sets are  parameterized by the torus of dimension $2d$,
\begin{equation}
(\R^d\times\R^d)/\widetilde L \simeq (\R/\Z)^d=: \T\,.
\end{equation}
We note,  though, that  the parametrization need not necessarily be $1-1$,
i.e. in general, $\Lambda_{(x,y)}=\Lambda_{(x',y')}\nRightarrow (x,y)=(x',y')\mod \widetilde L$.
In the sequel, for simplicity, we shall usually write $(x,y)_L$ for the congruence class
$(x,y)\mod \widetilde L.$

There is a natural measure, the Haar measure,  $\theta_\T$ on $\T$.
This measure is the obvious `area' measure in the case of $\T^2$,
and `length' measure for $\T^1$. It is invariant under the
$\R^d$-action: $\R^d$ acts on $\R^{2d}/\widetilde L$ by
\begin{equation}
t+(x,y)_L=(t+x,y)_L\,.
\end{equation}

Of particular importance to us is natural embedding (see Fig.~2.)
\begin{equation} \label{torusembedding}
\R^d \, \longrightarrow \, \T \quad \quad t \mapsto (t,0)_L  \,,
\end{equation}
 which lies behind the connection between almost periodicity in physical
 space and periodicity in some higher dimensional setting. The image of this mapping is easily
 established to be dense in $\T$.
 \smallskip

Now start with $\Lambda=\Lambda_{(0,0)}=\Lambda(\Omega)$,
and translate it by elements $t\in\R^d$:
\begin{equation}
t+\Lambda(\Omega)=t+\Lambda(0+\Omega)=\Lambda_{(t,0)}\,.
\end{equation}
Form the local hull $X(\Lambda)$, the closure of the set
of all translates $\Lambda_{(t,0)}$ of $\Lambda$ under the local topology \eqref{hull}.

\begin{prop} \cite{Schlott}
There is a continuous mapping,
\begin{equation}
\beta : X(\Lambda)\ \longrightarrow\ \T\,,
\end{equation}
 called the torus parametrization, such that
\begin{itemize}
\item[1)] $\beta$ is onto;

\item [2)] $\beta$ is 1 - 1 almost everywhere, seen from the perspective of Haar measure on~$\T$;

\item [3)]  for all $t\in\R^d$, for all $\Lambda'\in X(\Lambda)$,  one has $\beta(t+\Lambda')= t+\beta(\Lambda')$.

\item [4)] $\beta(t+\Lambda)=(t,0)_L$ for all $t\in\R^d$.      \qed
\end{itemize}
\end{prop}

$X(\Lambda)$ and $\T$ are both compact spaces with natural $\R^d$ action, and $\beta : X(\Lambda)\rightarrow\T$
is an onto $\R^d$ mapping. But $X(\Lambda)$ and $\T$ are subtly different. Although $t+\Lambda\in X(\Lambda)$
and $t+\Lambda\mapsto\Lambda_{(t,0)}$ for all $t$, not every element of $X(\Lambda)$ is a $\Lambda_{(x,y)}$
for some $(x,y)$. Rather, when  $(-y+\partial\Omega)\cap L'\neq\emptyset$, then for all $x\in\R^d$
there are always $\Lambda_1\neq\Lambda_2$ of elements of $X(\Lambda)$ that are mapped by $\beta$
to the same $(x,y)_L$. When $\tilde z=(z,z')\in \widetilde L$ with $z' \in(-y+\partial \Omega)$
then there will always be $\Lambda_1,\Lambda_2\in X(\Lambda)$ with $\beta(X(\Lambda_1))= \beta(X(\Lambda_2))=(x,y)_L$,
yet $x+z\in\Lambda_1,$  $x+z\notin\Lambda_2$. In other words, there are ambiguities regarding the lattice points
that project onto the boundary $\partial\Omega$ of $\Omega$.
An example of this is shown in the footnote appearing in
\S\ref{ssec_Fibonacci_model_set}. The fact that $\beta$ is $1-1$ almost everywhere is due to our assumption
that the boundary of $\Omega$ has measure 0.

When we say that $\beta$ is 1 - 1 almost everywhere, we mean that the set $A$ of points $\xi$ in $\T$,
for which there is more than one point set over $\xi$, satisfies $\theta_\T(A)=0$.

There is a {\it unique $\R^d$-invariant ergodic measure} $\mu$ on $X(\Lambda)$ with $\mu(X(\Lambda))=1$.
In fact $\beta$ relates $\mu$ and $\theta$:
\begin{equation}
\beta(\mu)=\theta\,.
\end{equation}
Or more specifically, $\theta(A)=\mu(\beta^{-1}A)$ for all measurable subsets $A$ of $\T$.
With $\mu$ in hand, we can introduce the space $L^2(X(\Lambda),\mu)$ of square integrable
functions on $X(\Lambda)$.  As already pointed out, the natural action of $\R^d$ on this is unitary.

\subsection{From hulls to tori}\label{lfoaps}\

Square integrable functions on $X(\Lambda)$ and square integrable functions on $\T$ can be identified,
\begin{equation}\label{identif}
L^2(X(\Lambda),\mu)\simeq L^2(\T,\theta)\,.
\end{equation}
The isomorphism is easy to understand:
\smallskip

\centerline{
\diagram
X(\Lambda) \rto^{\beta} \morphism\dashed\tip\notip[dr]_{G\circ\beta}& \T \dto^G\\
& \C
\enddiagram
}

\noindent gives us a map
\begin{gather*}
L^2(\T,\theta)\quad\longrightarrow\quad L^2(X(\Lambda),\mu).
\end{gather*}
Since $\beta$ is almost everywhere 1 - 1, the map is a bijection.
\smallskip

This allows us to analyze functions on  $X(\Lambda)$ by treating them as functions on~$\T$.
The advantage of this is that
functions in $L^2(\T,\theta)$ have Fourier expansions

\begin{equation}
\wt F(z)=\sum_{\wt k\in \wt{L}^\circ}a_k \,e^{2\pi i\l \wt k\mid z\r}\,,
    \qquad z\in\T,\  a_k\in\C\,.
\end{equation}
Here $\langle\cdot,\cdot\rangle$ is the natural dot product on $\R^d\times\R^d=\R^{2d}$ and
$\wt{L}^\circ$ is the lattice which is $\Z$-dual to $\wt L$:
\begin{gather*}
\wt k\in\wt{L}^\circ\Longleftrightarrow\langle\wt k\mid\wt x\rangle\in\Z\,,
\qquad\text{for all}\quad \wt x\in\wt L\,.
\end{gather*}
Assuming that the dot product is rational valued on $\wt{L}$, this new lattice is in the $\Q$ span of $\wt L$ and its
elements $\wt k$ have the same type of decompositions $\wt k = (k,k')$ as elements of
$\wt L$. In particular, for each $\wt k$ there is a unique $k \in \R^d$. We prepare for
the ultimate reduction of everything to the physical space by already using the symbols
$a_k$ rather than $a_{\wt k}$ for the coefficients of the Fourier expansion. We shall write $L^\circ$
for the projection of $\wt L^\circ$ on the physical side, so we can just as well write $k\in L^\circ$
as $\wt k\in \wt{L}^\circ$. We will call $L^\circ$ the dual (or
$\Z[\tau]$-dual) of $L$. The mapping $(\cdot)' : L^\circ \rightarrow \R^d$ here is compatible
with the one on $L$, indeed $\Q L^\circ = \Q L$.

The corresponding functions on $X(\Lambda)$ have similar expansions. This works as follows:
If $F:X(\Lambda) \longrightarrow\C$ corresponds to $\wt F : \T\longrightarrow \C\,$
then, for $\Lambda_1\in X(\Lambda)$ with $\beta(\Lambda_1)=(x,y)_L$, we have as $L^2$-functions
\begin{equation}
F(\Lambda_1)=\wt F((x,y)_L)
   =\sum_{k\in L^\circ}a_k \, e^{2\pi i\l \wt k\mid(x,y)\r}\,.
\end{equation}

\subsection{Local functions on model sets}\label{modelsets}\

Let $\Lambda$ be a model set arising from the cut and project scheme of \eqref{cps}.
Let $\T=\T^{2d}=(\R^d\times\R^d)/L$ be the torus with torus parametrization
\begin{gather*}
\beta : X(\Lambda)\ \longrightarrow\ \T\,.
\end{gather*}
Then we have the identification \eqref{identif} of the corresponding $L^2$
spaces. Each element $\Lambda'\in X(\Lambda)$ maps by $\beta$ to a point $\beta(\Lambda')$ in $\T$.
We also know $\beta(\Lambda)=(0,0)_L$, and $\beta(t+\Lambda)\rightarrow (t,0)_L$.
So we know how $\beta$ works on $\R^d+\Lambda$.

Suppose $f$ is a local function with respect to the model set $\Lambda$.
From the local function $f$ we have its extension $F\in L^2(X(\Lambda),\mu)$ which is continuous.
Then we obtain $\wt F\in L^2(\T,\theta)$,  where
\begin{equation} \label{basicFunctionRelation}
\wt F((t,0)_L)= \wt F(\beta(t+\Lambda))=F(t+\Lambda)= f(t)\,,
\end{equation}
and we can write
\begin{gather*}
\wt F(\cdot)=\sum_{k\in L^\circ}a_k\, e^{2\pi i\l \wt k\mid\cdot\r}\,,
\end{gather*}
where $L^\circ$ is the dual of $L$.
\begin{definition} The \emph{  Fourier-Bohr expansion} of the local function $f$ is
\begin{equation}\label{Fourier}
f(t)=F(t+\Lambda)=\wt F((t,0)_L)
   =\sum_{k\in L^\circ}a_k\,e^{2\pi i\l \wt k\mid(t,0)\r}
   =\sum_{k\in L^\circ}a_k\,e^{2\pi i\l k \mid t \r}\, .
\end{equation}
\end{definition}

\begin{figure}[ht]
\centering
\includegraphics[scale=0.5]{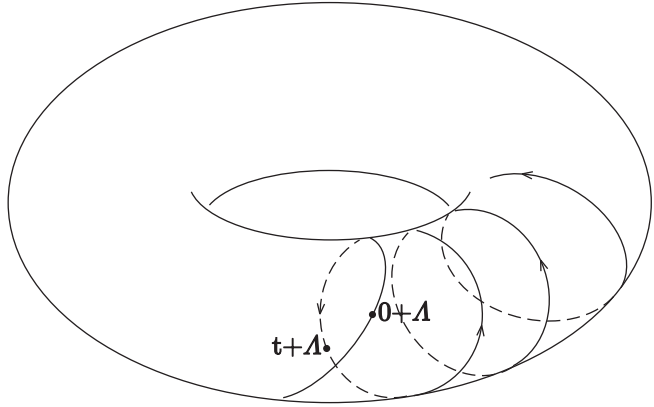}
\caption{A fragment of the orbit of $\Lambda$ as seen in the torus parametrization.}
\end{figure}

Our study of almost periodic functions on $\Lambda$ becomes the study of functions on $\T$
and their restrictions to the `spiral' orbit $\R^d+\Lambda$ in $\T$ given by the embedding (\ref{torusembedding}),
i.e. restriction to $(\R^d,0)_L$.

\subsection{Fourier coefficients}\label{fouriercoefficients}\

Let us continue with the situation in \S\ref{modelsets}. For all $x\in\T$,
\begin{align}\label{coeff}
\wt F(x)&=\sum_{k\in L^\circ}a_k\,e^{2\pi i\l \wt k\mid x\r}\\
\text{where}\qquad
a_k&=\int_\T e^{-2\pi i\l \wt k\mid x\r}\wt F(x)\,d\theta_{\T}(x) \, .
\end{align}
Unfortunately, we don't have total control over $\wt F$. We know it only on $(\R^d,0)_L$.
To compute $a_k$ out of $f$ alone, we use the Birkhoff ergodic theorem\footnote{It is hard to find
convenient references for the Birkhoff ergodic theorem in the
form we need it. Most references prove it over $\Z$ or $\R$. In \cite{Keller}
there is a proof over $\Z^d$ which is easy to adapt to $\R^d$.} : for all continuous
functions $\wt G$ on $\T$,
\begin{equation} \label{Birkhoff}
\int_\T \wt G(x)\,d\theta_{\T}(x)
  =\lim_{R\rightarrow\infty}
     \frac1{\text{vol}\,B_R}\int_{B_R} \wt G((t,0)_L)\,dt\,.
\end{equation}

Thus
\begin{align}
a_k&=\lim_{R\rightarrow\infty}\frac1{\text{vol}\, B_R}
       \int_{B_R}e^{-2\pi i\l \wt k \mid (t,0)\r}\wt F((t,0)_L)\,dt \nonumber \\
   &=\lim_{R\rightarrow\infty}\frac1{\text{vol}\,B_R}
       \int_{B_R}e^{-2\pi i\l k \mid t\r}f(t)\,dt\,. \label{coeff2}
\end{align}
Here we use  $F((t,0)_L)=f(t)$ and $\wt k=(k,k')$, so
\begin{gather*}
\l \wt k \mid (t,0)\r=\l k,t\r+\l k',0\r=\l k,t\r\,.
\end{gather*}

The averaging sequence $\{B_R\}$ that we have employed here can be replaced
by any unbounded ascending sequence $\{A_n\}$, where $A_n\subset\R^d$ is compact,
$\bigcup A_n = \R^d$, and the boundary of $A_n$ has measure 0 for all $n$.
In actual practice one should adapt the averaging sequence to the problem at hand.

\section{Discretization} \label{discretization}
\subsection{Main components}\label{Tdm}

Our objective is to devise a discrete method by which to estimate the coefficients $a_k$
of~\eqref{coeff2} of a local function $f$ with respect to a model
set $\Lambda = \Lambda (\Omega)$. This means replacing the integral by
a finite sum of values of the integrand.
The problem is to do this in such a way that it respects the cut and project scheme in which
the model set lives, can be guaranteed to converge in the limit to the required integral,
and can be carried out efficiently from a computational point of view.

There are three components to this:
\begin{itemize}
\item[(i)] deciding on a suitable domain of integration (what should we use for $B_R$?);
\item[(ii)] creating the points of evaluation of the integrand, including how many there should be;
\item[(iii)] deciding to which set of Fourier coefficients (which values of $k$) we should restrict our attention.
\end{itemize}

A key feature of discrete methods involving periodic functions is the use of finite groups
arising from refinements of the period lattice and quotients of its
dual lattice, e.g.  \cite{ MP06}. We need to translate this concept into the context
of cut and project schemes. The set of points on which the integrand is evaluated
is created out of the same cut and project process that creates the original model set.
The ingredients are a choice of a suitable lattice $\wt{L_N} \supset \wt L$, which then
gives rise to the finite group $\wt{L_N}/\wt L$. The data points in $\R^d$ at which
computations of our functions will be made come by projection into physical space
of a suitable set of coset representatives of $\wt{L_N}$ modulo $\wt L$.
The corresponding frequencies (wave vectors) $k$ are chosen from the dual
lattice $L^\circ$. The choice
of values of $k$ at which we should evaluate the Fourier coefficients $a_k$
come by selecting suitable representatives of $\wt L^\circ$ modulo $\wt{L_N}^\circ$.
The key point is the duality
\begin{gather*}
\langle \cdot\,, \cdot\rangle : \wt{L}^\circ/\wt{L_N^\circ} \times \wt L_N/\wt L \longrightarrow \frac{1}{N}\Z/\Z\,.
\end{gather*}

\subsection{Outline of the discretization process} \label{generalSetting}
In this section we give more precise details as to how the goals of \S\ref{Tdm} can be achieved.
The process necessarily involves
a number of decisions, which can only be made in the context of the situation at hand.
In \S\ref{Fibonacci} and \S\ref{TwoEE} we shall see how this looks in particular examples.

It should be noted that although getting the computational details set up is somewhat involved,
these details depend only on the cut and project scheme and the degree of accuracy required from the computation.
Once this  is established the data points
and choices of frequencies are already determined, and they suffice for the Fourier analysis
of \emph{all}  functions that arise out of the same almost periodic family, and are algorithmically easy to compute.

We begin with the cut and project scheme \eqref{cps} with torus $\T$ and note the
natural extension of the mapping $(\cdot)'$ to the rational span of the module $L$:
\begin{alignat}{3}\label{cpsmap}
\Q L\ &\overset{1-1}{\longleftarrow}\quad
      &&\Q\wt L&&\longrightarrow\quad\Q L'\\
 x\   &\tomaps
      &\widetilde{x}&=(x,x')
      &\quad&\mapsto\quad\quad x'
      \notag
\end{alignat}

We assume that $\R^{2d} \simeq \R^d\times\R^d$ is supplied with the standard
dot product (denoted $\langle \cdot | \cdot \rangle$), and then define the dual lattice:
\begin{gather*}
\wt{L}^\circ= \{Y\in\R^d\times\R^d\ :\ \langle Y\mid\wt{x} \rangle \in\Z\ \;
\text{for all}\ \wt{x}\in\wt{L}\}.
\end{gather*}

Then $\wt{L}^\circ$ is a $\Z$-module of the same rank as $\wt{L}$, namely $2d$. There
is a cut and project scheme dual to \eqref{cps} of which $\wt{L}^\circ$ is the lattice \cite{RVM}:
\begin{alignat}{3}\label{cpsdual}
\R^d\ &\overset{||}{\longleftarrow}\quad
      &\R^d
      &\times\R^d
      &
      &\quad\overset{\perp}{\longrightarrow}\qquad\R^d\notag
      \\
      &&&\cup
      &&
      \\
      L^\circ &\overset{1 - 1}{\longleftarrow}
      &&\widetilde{L}^\circ
      &&\overset{\text{dense image}}{\longrightarrow}\quad (L^\circ) '\, .
      \notag
\end{alignat}
We shall use the same type of notation as in \eqref{cpsmap} for this scheme too.
It arises by taking the Pontryagin duals of all the groups in \eqref{cpsmap}, whereupon
$\wt{L}^\circ$ appears as the dual of the torus $\T$. For more on dual cut and project schemes see \cite{RVM}.

Dualizing can be considerably simplified if the inner product on
$\R^d \times \R^d$ is rational valued on the lattice $\widetilde L$. This quite often
happens in actual practice. For instance, it happens in the Fibonacci example below,
where the inner product arises from the trace form on $\Q[\sqrt 5]$.
When this happens one can identify $\wt{L}^\circ$
as a subset of $\Q\wt{L}$ and thereby avoid having to find the new $(\cdot)'$ mapping.
However, in the general situation such simplifications
need not exist,  and we do not assume them here.
 \medskip

In reading what follows it is good to keep in mind the equation
\begin{equation}\label{expEq}
 e^{2 \pi i \langle \tilde k| \tilde s\rangle} = e^{2 \pi i \langle k| s\rangle}
e^{2 \pi i \langle k'| s'\rangle}
\end{equation}
which lies at the bottom of the approximation.
The discrete Fourier analysis is accomplished by a dual pair of finite
groups from which the $\tilde k$ and the $\tilde s$ will come.
The actual values of $\tilde k$ and $\tilde s$ are important only modulo
the lattices $\wt {L_N}^\circ$ and $\wt L$ respectively, and this freedom lies
at the heart of the process.

The values of $s$ should be in the range $A$ of our integration, and ideally
we would have the corresponding $s' =0$ since the $\tilde s$ are supposed
to be representing points of the physical space $\R^d$.
However, the latter is not possible, so we attempt to choose the $s \in A$
along with $s'$ as small as possible. The approximation then works by throwing away the term which involves $k',s'$  in
\eqref{expEq}. The set of values of
$\tilde k$ is constrained primarily by the requirement that the values of $|k|$
should be small, see the discussion after Step 5 below.

\subsection{The six steps}\label{6steps}
{\bf Step 1}: Choose a finite subgroup of $\T$ of order $N$. This appears in the form
$\widetilde{L_N}/\widetilde{L}$ where $\widetilde{L_N}  \supset \widetilde L$ is a lattice refining $\widetilde L$.
The number $N$ will determine the number of points of evaluation in approximating the integrals by sums.
We have $\wt{L}_N\subset\Q\wt{L}$ and it has a $\Z$-dual $\wt{L}_N^\circ\subset\wt{L}^\circ$
which is of index $N$ in $\wt{L}^\circ$.

This affords the natural pairing
\begin{equation}
 \wt{L}^\circ/\wt{L}_N^\circ\times\wt{L}_N/\wt{L}
\quad\longrightarrow\quad\frac{1}{N}\Z/\Z
\end{equation}
induced by $\langle \cdot\mid\cdot \rangle $ (and still denoted by $\langle \cdot\mid\cdot \rangle $).
Using the $(\cdot)'$ mappings, we obtain, in the obvious notation,  the $\Z$-modules
$L_N \supset L$ , $L_N^\circ\subset L^\circ$, $\wt{L}_N/\wt{L}\simeq L_N/L$,
$\wt{L}^\circ/\wt{L}_N^\circ\simeq\wt{L}/\wt{L}_N$, and an induced pairing
\begin{gather*}
\langle \cdot\mid\cdot\rangle \ :\ L^\circ/L_N^\circ \times L_N/L
   \quad\longrightarrow\quad\frac{1}{N}\Z/\Z\,.
\end{gather*}

Replacing $\wt L$ by $\wt L_N$, we have the refined cut and project
scheme

\begin{equation}\label{cpsrefined}
\begin{alignedat}{5}
&\R^d&&\quad\longleftarrow\quad&\R^d&\times\R^d&&\quad\longrightarrow\quad&&\R^d \\
&&&\quad&&\cup&&\quad&& \\
&L_N&&\quad\overset{1-1}\longleftrightarrow\quad&&\wt{L}_N
         &&\quad\longleftrightarrow\quad&&L_N' \\
\end{alignedat}
\end{equation}
and similarly its dual.

\medskip
\noindent
{\bf Step 2}: Choose a fundamental domain $C$ for $\wt L$.
A canonical choice would be the Voronoi cell of $\widetilde L$ at 0, but any other choice is allowable. In the
examples below we use the parallelogram defined by a pair of basis
vectors of $\wt L$. Cover $(\R^d, 0) \subset \R^{2d}$  with a set of translates $\tilde t + C$ of $C$
by elements of $\wt L$.  The projection of these cells into the physical space $\R^d$ covers it,
though in general the projected $\tilde t + C$ have many overlaps.

\medskip
\noindent
{\bf Step 3}: Form $\wt S := \widetilde{L_N} \cap C$. This provides a complete set
of representatives in $\widetilde{L_N}$ for the group $\wt{L}_N/\wt{L}$.

\medskip
\noindent
{\bf Step 4}: Choose a region $A$ which will be the delimiting the range
over which the Fourier coefficients of $f$ will be estimated in the form
\begin{gather*}
a_k=a_{\wt k} \simeq \frac1{\text{vol} \,A}
\int_{A}e^{-2\pi i \langle k\mid t \rangle}f(t)\,dt.
\end{gather*}
These integrals have to be computed for values
of $k$ that come from $L^\circ/L^\circ_N$.

Take as $A$ the image of a finite set of the translates of
$C$ appearing in Step 2, i.e.
\begin{gather*}
A = (\bigcup_{\tilde t \in \wt T} \tilde t + C)^{||}
\end{gather*}
for some finite subset of $\wt T$ elements of $\wt L$.
The choice of $A$ is again determined by the problem at hand.

\smallskip

We need next to determine a set of data points in $\R^d$ which will serve to replace
the integrals of Step 4 by finite sums. This is the purpose of the next step.

As we pointed out above, we are free to translate  the elements of $\wt S$ by $\wt L$ as we please,
and we wish to do this so that the projected images are in our region of integration
$A = \bigcup_{\tilde t \in \wt T} (\tilde t + C)^{||}$.
We also keep in mind that we wish to do this so as to minimize the size of the corresponding $s'$.

\medskip
\noindent
{\bf Step 5}:  For each $\tilde s = (s,s') \in \wt S$ find a $\wt{t(s)} \in \wt T$ for which
$|s' +t(s)'|$ is minimal. Then the set of data points is
\begin{gather*}
D:= \{ s + t(s) \,: \,  \wt s \in  \wt S \}.
\end{gather*}

At this point, for each $k\in L^\circ$ we have
\begin{gather*}
a_k:= \frac{1}{{\rm vol} \,A} \sum_{u\in D} e^{-2\pi i \langle k|u\rangle} f(u)
\end{gather*}
and the resulting approximation of $f$ is
\begin{gather*}
f(x) \simeq \sum_{k\in K} a_k e^{2\pi i \langle k|x\rangle}.
\end{gather*}
The set $K$ is to run over a complete set of representatives of $L^\circ/L_N^\circ$.
The choice seems free, but one may assume that in most cases the lower frequencies (smaller
$|k|$) are most essential in approximating $f$ by using only finitely many of its
frequencies. For this reason we have:

{\bf Step 6:}  Choose $k$ for each class of $L^\circ/L_N^\circ$ with
$|k|$ taken as small as possible.

 This concludes the broad description of the algorithm.

\section{A Fibonacci example}\label{Fibonacci}

To make all this more concrete we work through the details of a one dimensional example,
the well-known Fibonacci sequence, where the cut and project scheme lives in two dimensions
and the geometry of the data set $D$ and the
set of translates $\tT$ via a refinement lattice and new windows are easily visualized.
This involves first setting up
the cut and project scheme in detail, \S\ref{Fcps} and \S\ref{dualLattice}, and then describing a Fibonacci
point set arising from the standard Fibonacci substitution in terms of~it,~\S\ref{ssec_Fibonacci_model_set}.
We then follow Steps 1 through 6 of
\S\ref{6steps}, which provide the data points and corresponding frequencies
that will work for the analysis of any local function that we may choose.
In \S\ref{TwoEE} we apply this information to two simple examples of local functions
to see how well the methods actually work.

One of the great virtues of the cut and project method is that it is primarily
an algebraic tool and does not require great geometric insight to use it.
Given that for aperiodic structures in dimension greater than $1$ we almost always
in the situation of lattices of rank greater than $3$, and hence forced into spaces
in dimensions greater than $3$, this type of algebraic formalism is of enormous value. However, in the
Fibonacci example, in which everything can be done in $2$-dimensions, it is useful
to see the underlying geometry explicitly. Thus we have gone to some effort to show the geometric meaning of the central
feature of the method, that is, the creation of the data points, and to show what the
approximations look like and how good they are in comparison with
exact computation.
In actual practice all this is unnecessary. The only part of the algorithm that requires any serious
insight into the geometry (and it is actually trivial in the
Fibonacci example) is the selection of the fundamental domain and the
translates of it that are to be used. It is their projection that makes the domain in physical
space in which the data will lie.

\subsection{The Fibonacci cut and project scheme}\label{Fcps}

Let $\tau:=\tfrac12(1+\sqrt5)$ and let $Z[\tau]=\Z+\Z\tau$. Then $Z[\tau]$ is the ring of integers
of the field $\Q[\tau]=\Q[\sqrt5]$ and $\tau^2=\tau+1$. Let $(\cdot)'$ on $Z[\tau]$ and $\Q[\tau]$
be the conjugation that interchanges $\sqrt5$ and~$-\sqrt5$.

We define
\begin{gather*}
\wt{\Z[\tau]}:=
     \{(x,x')\ :\ x\in\Z[\tau]\}\subset\R\times\R\,.
\end{gather*}
$\wt{\Z[\tau]}$ is a lattice in $\R\times\R$ and its natural projections
\begin{gather*}
\R\tomaps\R\times\R\mapsto\R
\end{gather*}
into $\R$ provide the set-up for the Fibonacci cut-and-project scheme:

\begin{equation}\label{cpscheme}
\begin{alignedat}{5}
&\R&&\longleftarrow\quad&\R&\times\R&&\longrightarrow\quad&&\R \\
&\cup&&\quad&&\cup&&\quad&&\cup \\
&\Z[\tau]&&\quad\longleftarrow\quad&&\wt{\Z[\tau]}
         &&\quad\longrightarrow\quad&&\Z[\tau] \\
&x       &&\quad\tomaps\quad&\wt x&=(x,x')
         &&\quad\mapsto\quad&
         &x'\,.
\end{alignedat}
\end{equation}
A natural basis of $\wt{\Z[\tau]}$ is $\{(1,1)$, $(\tau,\tau')\}$
and the standard inner product on $\wt{\Z[\tau]}$ is defined by
\begin{equation}
(\wt x\mid\wt y)=(2x\cdot y)_\Z= \delta xy + (\delta xy)'\,,
\end{equation}
Here the notation $(\ \cdot\ )_\Z$ indicates taking the rational component $a$ of
$2x\cdot y\in\Z[\tau] =:a + b\tau$, and
\begin{align}\label{delta}
\delta&=(\tau\sqrt5)^{-1} =(\tau(\tau-\tau'))^{-1}=\tfrac1{\tau^2+1}\,,\nonumber
\\
\delta'&=(-\tau'\sqrt5)^{-1} =(\tau'(\tau'-\tau))^{-1}=\tfrac1{\tau'^2+1}\,,
\end{align}
In particular,
\begin{gather*}
(\wt 1\mid\wt 1)=(2(1\cdot1))_\Z=2\,,\quad
(\wt\tau\mid\wt\tau)=(2\tau^2)_\Z=2\,,\quad
(\wt 1\mid\wt\tau)=(2\tau)_\Z=0\,.
\end{gather*}
The geometry of the fundamental cell for the lattice $\wt {\Z[\tau]}$ is illustrated in
Fig.~3. More details of the material here can be found in \cite{CMP}.

\subsection{The dual lattice}\label{dualLattice}

The inner product $(\cdot\mid\cdot)$ allows us to identify $\wt{\Z[\tau]}^\circ$
inside the rational span of $\wt{\Z[\tau]}$ and to express $\l\cdot\mid\cdot\r$
in terms of $(\cdot\mid\cdot)$. The basis dual to $\{(1,1),(\tau,\tau')\}$ is given by
\begin{equation}
\wt\omega_1=\tfrac12\wt1=(\tfrac12,\tfrac12)\,,
\qquad
\wt\omega_2=\tfrac12\wt\tau
           =(\tfrac{\tau}2,\tfrac{\tau'}2)\,,
\end{equation}
and the dual lattice is
\begin{equation}
({\wt{\Z[\tau]})^\circ}=\Z\wt\omega_1+\Z\wt\omega_2
                   =\tfrac12\wt{\Z[\tau]}\,.
\end{equation}
The elements of $({\wt{\Z[\tau]})^\circ}$ are always of the form $\wt k=(k,k')$ and we can write
\begin{equation}
\Z[\tau]^\circ:=\{k\ :\ \wt k=(k,k')\in({\wt{\Z[\tau]})^\circ}\}
\end{equation}
Then
\begin{equation*}
(\wt{\Z[\tau]})^\circ= \wt{\Z[\tau]^\circ}=
    \{\wt k\ :\ k=(k,k')\in\Z[\tau]^\circ\}\,,
\end{equation*}
and $\l\wt k\mid\wt x\r$ becomes $(\wt k\mid\wt x)$, which is more useful notation for the sequel.

The 2-torus $\T$ of the cut-and-project scheme \eqref{cpscheme} is then
\begin{gather*}
\T=(\R\times\R)/\wt{\Z[\tau]}\,.
\end{gather*}
Fourier series on $\T$ are expressed in terms of the characters $\chi_{\wt k}$, namely
\begin{equation}
\chi_{\wt k}({\wt x})= e^{2\pi i(\wt k\mid\wt x)}\,,
\quad\text{where}\quad
\wt k=(k,k')\in\Z\wt\omega_1+\Z\wt\omega_2\,.
\end{equation}
Here $\wt x$ can be arbitrary in $\R\times\R$, but we shall need to compute
only with $\wt x\in\Q\wt\omega_1+\Q\wt\omega_2$ for which $\wt\ $ is well-defined \eqref{cpscheme}.
For these elements $(\wt k\mid\wt x)=(2k\cdot x)_\Q$.

The product $(\cdot\mid\cdot)$ extends to $\R\times\R$,
but care has to be taken. For a typical element $a\wt1+b\wt\tau$, $a,b\in\R$ of our superspace  $\R\times\R$
the inner product is calculated as
\begin{gather*}
((a\wt1+b\wt\tau)\mid(c\wt1+d\wt\tau))=2ac+2bd\,,
\end{gather*}
where
\begin{gather*}
a\wt1+b\wt\tau=a(1,1)+b(\tau,\tau')
    =(a+b\tau,a+b\tau')\in\R\times\R\,.
\end{gather*}

Consider
\begin{gather*}
\wt k=a\wt\omega_1+b\wt\omega_2
     =\tfrac a2\wt1+\tfrac b2\wt\tau
     =\left(\tfrac{a+b\tau}2,\tfrac{a+b\tau'}2\right)
     =(k,k')\in\wt{\Z[\tau]}^\circ\,,
\end{gather*}
where $a,b\in\Z$.
Suppose we want to compute $(\wt k\mid(t,0))$, $t\in\R$. From
\begin{alignat*}{2}
(1,0) &=\tfrac1{\tau\sqrt5}\wt1+\tfrac1{\sqrt5}\wt\tau
     &&= \delta\wt1+\tau\delta\wt\tau\,,\\
(0,1) &=\tau^2\delta\wt1-\tau\delta\wt\tau
     &&=\delta'\wt1+\tau'\delta'\wt\tau\,,
\end{alignat*}
where $\delta,\ \delta'$ are from \eqref{delta}, we have
\begin{gather*}
(\wt k\mid(1,0))=2\delta k\,,
\end{gather*}
and so,
\begin{gather*}
(\wt k\mid(t,0))=2\delta kt\,,
   \qquad\text{for all}\quad t\in\R.
\end{gather*}
We are interested in model sets $\Lambda$ coming from the cut and project scheme \eqref{cpscheme}.
In the context of \eqref{cpscheme}, the Fourier-Bohr expansion \eqref{Fourier} assumes the simpler form
\begin{equation}
  f(t)=F(t+\Lambda)=\wt F((t,0)_{\wt{\Z[\tau]}})\\
            =\sum_{k\in\Z[\tau]^\circ}
            a_k \, e^{2\pi i2\delta kt}\,.
\end{equation}
We shall use the $a_k$ computed in the form
\begin{gather*}
a_k = \lim_{R\to\infty} \frac{1}{R}\int_0^R e^{-2 \pi i 2 kt} \,f(t) dt.
\end{gather*}

For future use note that
\begin{align}
(\wt k\mid (0,1)) &=
((\tfrac{a+b\tau}2,\tfrac{a+b\tau'}2)
      \mid \delta'\wt1+\tau'\delta'\wt\tau)
    =2k'\delta'  \notag\\
(\wt k\mid(0,u))  &=2uk'\delta'\,.
\end{align}

 \subsection{The Fibonacci model set}\label{ssec_Fibonacci_model_set}\

The standard $2$-letter Fibonacci sequence is the fixed point of the substitution $a \rightarrow ab$,
$b\rightarrow a$: namely,
\begin{gather*}
abaababa \dots\ .
\end{gather*}
With tile lengths $\tau$ for $a$ symbols and $1$ for $b$ symbols, and starting at
$0$, we get the sequence of tiles that cover the non-negative part of the real line.
The left-hand ends of these tiles,
\begin{gather*}
0,\,\tau,\,\tau+1,\,2\tau+1,\,3\tau+1,\,3\tau+2,\,4\tau+2,\,4\tau+3,\,
5\tau+3,\,\dots
\end{gather*}
form an infinite sequence of points on the non-negative real line.
This set appears explicitly as the non-negative part of the model set
\begin{gather*}
\Lambda = \Lambda([-1, \tfrac1\tau))
\end{gather*}
arising from the cut and project scheme \eqref{cpscheme}.\footnote{Alternatively
one can use $\Lambda((-1, \frac{1}{\tau}])$, which differs from $\Lambda$
in the two points coming from the ends of the interval $[-1, \frac{1}{\tau})$.
This is an example of a pair of sets that map to the same place in $\T$. This ambiguity shows up in
an interesting way later on, see \S\ref{sos}. } \label{TorusConfusion}
$\Lambda$ decomposes as $\Lambda_a = \Lambda([-\frac{1}{\tau^2}, \frac{1}{\tau}))$
and $\Lambda_b = \Lambda([-1, -\frac{1}{\tau^2}))$, which give the left-hand end points of
the $a$ and $b$ points respectively.

\begin{figure}[ht]
\centering
\includegraphics[scale=0.8]{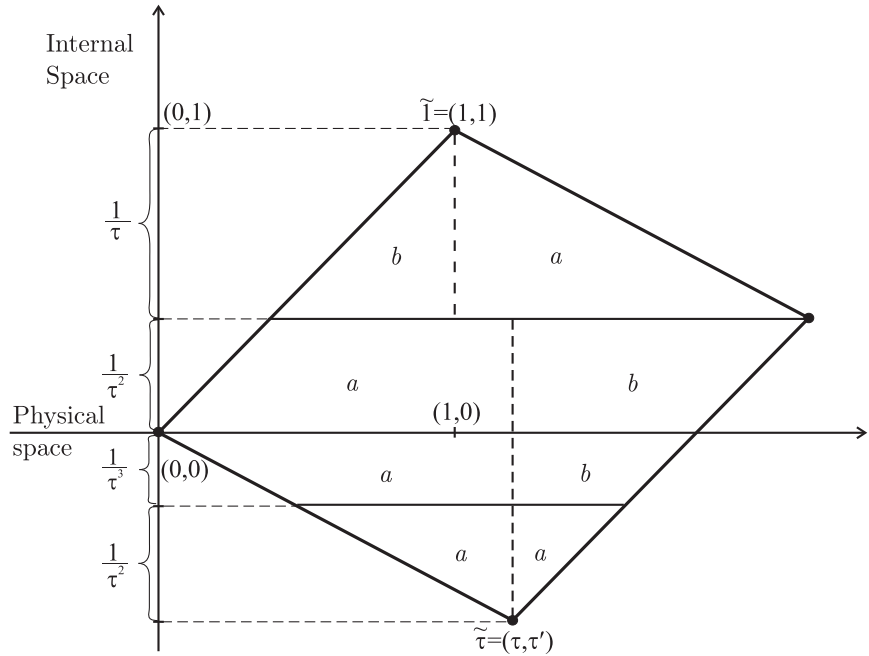}
\caption{The fundamental cell $C$ of the lattice $\wt{\Z[\tau]}=\Z(1,1)+\Z(\tau,\tau')$
is shown in terms of the standard coordinate system for $\R\times\R$.
When $C$ is imagined as a torus by identifying opposite sides and
the physical space is wrapped onto it using these identifications,
then the $a$ and $b$ regions shown here indicate which parts of the
physical space are lying in $a$ or $b$ tiles respectively. }
\end{figure}

A useful and commonly used way to visualize the distribution of $a$
and $b$ points is to view them on the torus $\T$ after the embedding \eqref{torusembedding}
of $\R$ in $\T$. The fundamental cell, see Fig.~3, with opposite edges identified is $\T$.
To see $\Lambda$ itself being formed, we start at $(0,0)$ and trace out $(t,0)$, $t\in \R$,
with the usual rules for exiting and re-entering the fundamental cell. We move continuously.
Our moving point has three `whiskers' attached to it.

\noindent
$a(1)$: a whisker of length $\tfrac1\tau$ facing vertically upwards;

\noindent
$a(2)$: a whisker of length $\tfrac1{\tau^2}$ facing down;

\noindent
$b(1)$: a whisker of length $\tfrac1\tau$ that faces down, but has an initial gap of size $\tfrac1{\tau^2}$.

The rule is this. As our point moves along the line on the fundamental cell,
if the $a(1)$ whisker hits $(1,1)$ (i.e. $(1,1)$ is close enough to get cut by this whisker),
we get an $a$-point of $\Lambda$. If the $a(2)$ whisker hits the point $(\tau,\tau')$,
we also get an $a$-point of $\Lambda$. If the $b(1)$ whisker hits the point $(\tau,\tau')$, we get a $b$-point.

Essentially the window, represented by the whiskers, is carried along with the moving point,
and exits and re-enters the fundamental cell with the moving point.

The advantage of this point of view is that it allows us to divide the fundamental
cell into regions which deliver the $a$- and $b$-points,
and thus to see what our function $f(t)$ looks like on the fundamental cell.

In Fig.~3 the vertical dotted lines are the key thing.
Our moving point moves to the right and every time it crosses a dotted vertical line,
we get a new point of $\Lambda$, thus starting a new interval.
We stay on that interval until the next crossing.

The $a$ and $b$ regions are indicated  in the Fig.~3.

\subsection{Discretization}\label{ssec_discr_of_analysis}\

We now go into the details of how to deal with the Fourier analysis using
discrete methods. Here we follow the six steps outlined in \S\ref{6steps}.

Write $\wt L$ for $\wt{\Z[\tau]}$, so $\wt L=\Z\wt1+\Z\wt\tau$.
Let $C:=\{u\wt1+v\wt\tau\ :\ 0\leq u,v>1\}$ be a fundamental region for $\wt L$.
Its volume is equal to $\sqrt5$ .

Fix any $N\in\Z_+$. We want to create a lattice $\wt {L_N}$ that is
a refinement of $\wt L$. In the \S\ref{6steps} we took $\wt{L_N}\supset \wt L$ with
$[\wt{L_N}:\wt L]=N$. In our present situation we use the most obvious lattices,
namely $\frac{1}{N} L$. These actually have index $N^2$, so there are some slight notational differences between
this section and \S\ref{6steps}.  Since $N\wt{L_N}\subseteq \wt L$, so $\wt{L_N} \subset \Q\wt1+\Q\wt\tau$.

Let $\tilde S := L_N \cap C$, so $\tilde S$ is a complete set of representatives
of $L_N \mod L$.

Together this completes steps 1, 2, 3 of \S\ref{generalSetting}.

For understanding the approximations better we also introduce
a fundamental region $C_N$ for $L_N$, chosen so that
\begin{equation} \label{newFD}
C = \bigcup_{\tilde s \in \tilde S} \tilde s + C_N \, .
\end{equation}

\smallskip

For Step 4 we choose as our delimiting range in $\R$ an interval
$[0,R]$ where $R$ is a positive real number taken so that
$(R,0)$ taken modulo $\wt L$ is on the boundary of $C$. Following along
the path $(t,0)$, $t\in \R_{>0}$, and wrapping around $C$ as indicated
in Fig.~4, this amounts to stopping at some point $R$ where the path is
just exiting $C$, so that we have an exact number of passes of $C$.
Thus the path $(t,0)$, $0\le t \le R$ involves an explicit set of translates
$\tt_i +C$, $i=1, \dots, M$ of $C$. Let $\tT = \{\tt_1, \dots, \tt_M\}$.

Let $f:\R\rightarrow\C$ be any continuous local function with respect to the model set
$\Lambda=\Lambda([-1,\tfrac1\tau))$, and let $\wt F:\T \rightarrow\C$
be its extension to a continuous function on $\T=R^2/\wt L$.
Generally we are interested in Fourier decomposition of $\wt F$
and along with it the corresponding decomposition of $f$. Thus we wish to compute things like
\begin{gather*}
\int_\T \wt F(x)\, e^{-2\pi i(\wt k\mid x)}d\theta_{\T}(x)\,.
\end{gather*}
Here $\wt F(\cdot)e^{-2\pi i(\wt k\mid\cdot)}$ is just some other continuous
function on $\T$ that is local with respect to $\Lambda$.
So it suffices to deal with some general continuous function $\wt G$
on $\T$ and its restriction $g(t)=\wt G((t,0))$ to the line $(\R,0)\mod \wt L$.

Let
\begin{gather*}
\varepsilon_N:=
    \sup_{i=1,\dots,N^2}\ \sup_{x\in \, \ts_i+C_N}|\wt G(x)-\wt G(\ts_i)|\,.
\end{gather*}
Our first estimate is
\begin{gather*}
\int_\T \wt G  d\theta_\T \; \sim \; \frac{\sqrt{5}}{N^2} \sum_{i=1}^{N^2} \wt G(\ts_i).
\end{gather*}
Since $\text{vol}\,C_{N}=\tfrac{\sqrt5}{N^2}$, the error in this is estimated by
\begin{align}\label{estim}
  \begin{aligned}
\left|\int_\T \wt G\,d\theta-\frac{\sqrt{5}}{{N^2}}\sum_{i=1}^{N^2}\wt G(\ts_i)\right|
&=\left|\sum_{i=1}^{N^2}\int_{\ts_i+C_{N^2}}
\left(\wt G(x)- \wt G(\ts_i)\right)\,d\theta(x)\right|\\
\leq\sum_{i=1}^{N^2}\int_{\ts_i+C_{N^2}}\varepsilon_N\,d\theta
&=\sum_{i=1}^{N^2}\varepsilon_N \,\rm{vol}\,C_{N^2}
  =\sqrt5\,\varepsilon_N\,,
  \end{aligned}
\end{align}
i.e. the error in our approximation is bounded by $\sqrt5\,\varepsilon_N$.
In practice $L_N$ and $N$ would be chosen so as to provide a suitably \emph{a priori} assigned value of $\varepsilon_N$.

\subsection{Restriction to $f$}\label{ssec_restric_to_f}\

In order to be useful, the computation must be restricted to values of the function $f$,
since this is all that one is given in practice. We wish to use an approximation of the form
\begin{gather*}
\frac1R\int_0^R g(t)\,dt \; = \;\frac1R\int_0^R \wt G(t,0)\,dt \; \sim \;  \int_\T \wt G  d\theta_\T \,,
\end{gather*}
in accordance with \eqref{Birkhoff}.

In this section we indicate the geometry behind Step~5.  Again, it should be pointed out that
in the final analysis much of the detail that appears here need not appear at all in the actual algorithm.

The path $\{(t,0)_L\mid 0\leq t\leq R\}$, wrapped around $C$, is divided by $C$ into segments $l_1,\dots,l_M$.

\begin{figure}[ht]
\centering
\includegraphics[scale=0.7]{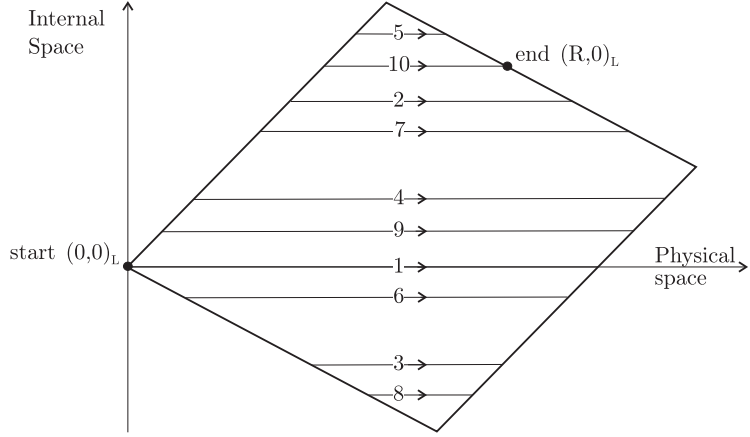}
\caption{Ten crossings of $C$ by the path $\{(t,0)_L\mid 0\leq t\leq R\}$ are shown in the torus.
(Opposite sides of the parallelogram coincide.)}
\end{figure}
Let $b_1<b_2<\cdots<b_M$ be the projections onto internal space of the boundary cutting points
of $\{(t,0)\ :\ 0\leq t\leq R\}$.
 Define
 \begin{gather*}
 c_0=\tau',\ c_1=\tfrac{b_1+b_2}2,\dots,\
 c_{M-1}=\tfrac{b_{M-1}+b_M}2,\ c_M=1\,.
 \end{gather*}
 The strips $S_i$ formed by passing the interval $[c_{i-1},c_i)$ in internal space through $C$
 in the direction of the physical axis, form a partition of $C$.
 Through each strip in $C$ runs a part $l_i$ of our line $\{(t,0)_L:0\leq t\leq R\}$.
 The idea is to use the intervals $[c_{i-1},c_i)$ as windows and the line segments $l_i$ as (part of) the
 physical space for a model set construction based on the lattice $\wt{L_N}$.
 This will then produce the points on $l_i$ which will be our data points for
 the evaluation of the functions $\wt G$, and then $g$.
 In other words, we are implicitly using the partial model sets $\Lambda_{\wt{L_N}}(S_i) \cap l_i$.

 Let $m(R):=\max\{\left|c_i-c_{i-1}\right|\}$ and
 \begin{gather*}
 \varepsilon'_N:=
 \sup_i\sup_{(x,v),(x,u)\in S_i}
        \left|\wt G(x,v)-\wt G(x,u)\right|\,,\qquad i=1,\dots,M\,.
 \end{gather*}

\begin{figure}[ht]
\centering
\includegraphics[scale=0.7]{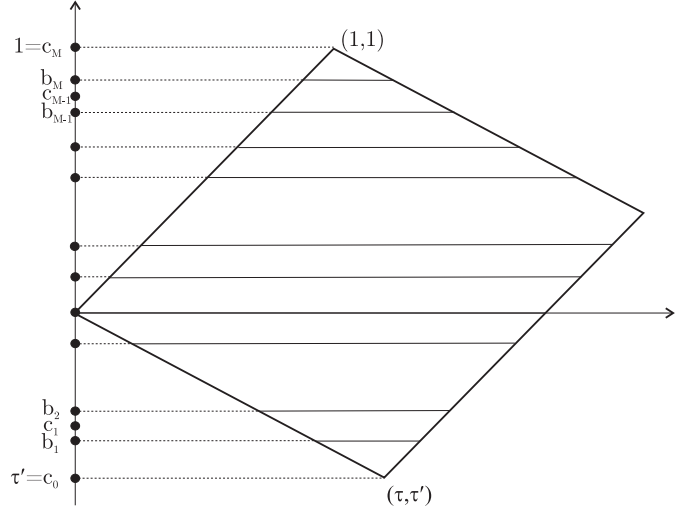}
\caption{The projections $b_1,\dots,b_M$ into internal space of the boundary
cutting points of the path. The points $c_0,\dots,c_M$ mark the
boundaries of the smaller windows into which the original window $[\tau',1)$ is partitioned.}
\end{figure}

Each $\tilde s_j \in \wt{S}$ lies in exactly one strip $S_i$. Let $(p_j,q_j)$
be its projection onto the line segment $l_i$. Then $(p_j,q_j)\equiv (u_j,0)_L$ for some $u_j\in[0,R]$.
Thus we obtain $\{(u_1,0),\dots, (u_{N^2},0)\}$ on our path $\{(t,0)_L\,:\,0\leq t\leq R\}$.

\begin{figure}[ht]
\centering
\includegraphics[scale=0.92]{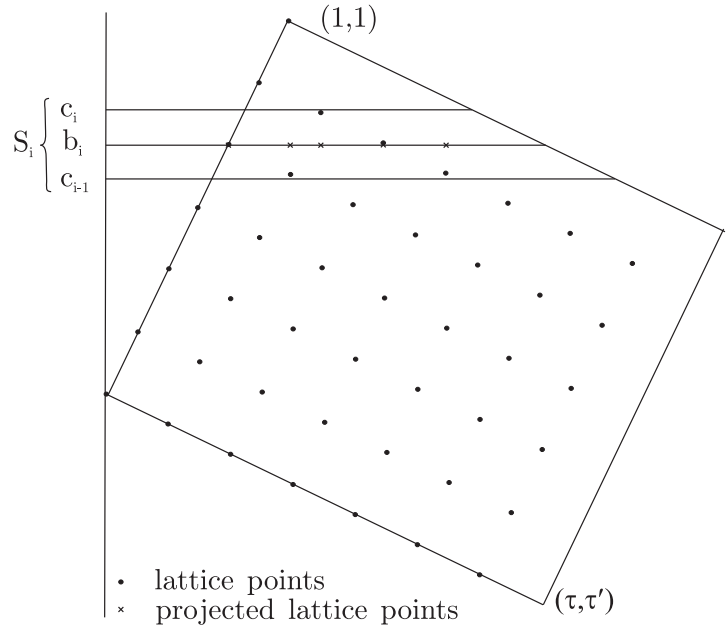}
\caption{The points $\tilde s_j$ lying in the strip $S_i$ formed by the window $[c_{i-1}, c_i)$
are projected onto the segment $l_i$ of the path $\{ (t,0)_L : 0\le t \le R\}$ producing data points
$u_j$, shown here as small crosses.}
\end{figure}

We have the estimate \eqref{estim}. Moreover,
\begin{gather*}
\frac{\sqrt{5}}{{N^2}}\sum_{i=1}^{N^2} \wt G(\tilde s_i)
 =\frac{\sqrt{5}}{{N^2}}\sum_{i=1}^{N^2}( \wt G(\tilde s_i)- \wt G(p_i,q_i))
 +\frac{\sqrt{5}}{{N^2}}\sum_{i=1}^{N^2} g(u_i)\,,
\end{gather*}
since
\begin{gather*}
\wt G(p_i,q_i)= \wt G((u_i,0))=g(u_i)\,.
\end{gather*}
Thus
\begin{equation} \label{stripSum}
\left|\frac{\sqrt{5}}{{N^2}} \sum_{i=1}^{N^2} \wt G(\tilde s_i)
 -\frac{\sqrt{5}}{{N^2}}\sum_{i=1}^{N^2} g(u_i)\right|
 \leq \frac{\sqrt{5}}{{N^2}}\sum_{i=1}^{N^2}\varepsilon'_N = \sqrt{5}\, \varepsilon'_N \,,
\end{equation}
since $\tilde s_i$ and $(p_i,q_i)$ both have second component in the same interval $[c_{i-1},c_i)$.

Combining \eqref{estim} and \eqref{stripSum}, we have
\begin{equation}
\left|\int_\T \wt G\,d\theta
 -\frac{\sqrt{5}}{{N^2}}\sum_{i=1}^{N^2} g(u_i)\right|
 <\sqrt5\,(\varepsilon_N+\varepsilon'_N)\,.
\end{equation}

This provides a method of estimating the integral $\int_\T \wt G\,d\theta$ using only $g$
on $[0,R]$ along with well chosen points in the interval. The two parameters $N$ and $R$ control the estimates.

In spite of the apparent complexity of strips, what is going on is easy to implement.
For each $\tilde s_i = (s_i,s'_i)$ there is a translation vector $\tt_j \in \tT$ for which
$|s'_i + t'_j|$ is minimal. The corresonding data point is then
$u_i := s_i + t_j$. This is Step 5 of \S\ref{generalSetting}.

\section{Two explicit examples}\label{TwoEE}

 \subsection{Computation of Fourier coefficients}\label{CFourier}

In this section we apply the methods outlined above to two almost periodic functions
$f:\R\rightarrow\R$, both local with respect to the model set
$\Lambda=\Lambda([-1,\tfrac1\tau))$ of \S\ref{ssec_Fibonacci_model_set}.
The~first is the distance-to-the-nearest neighbour function
\begin{equation} \label{testfunction}
f(t) = \textrm{the distance of $t$ to the nearest point of $\Lambda$} \, .
\end{equation}
This is a continuous and piecewise linear function that is local with respect to
$\Lambda$.

The second is the function
\begin{equation}\label{testfunction2}
f: \quad f(x) = \begin{cases} 1, &\text{if $x$ is in a long interval}\\
-1, &\text{if $x$ is in a short interval} \,,
\end{cases}
\end{equation}
which local, but only piecewise continuous with breaks wherever $x$ switches from
a long to a short interval.

Graphs of these functions are shown (solid lines) in Fig. 8 and Fig. 13 respectively.

Our objective is to see how well the approximations we have discussed compare
with the actual functions when the calculations are done
with specific choices of data points.

We approximate each of the two functions  by a finite number of terms of its Fourier-Bohr
expansion \eqref{Fourier}, which in our setting here reads:
\begin{gather} \label{FBexp}
f(x)=\sum_{\tilde k\in \widetilde{\Z[\tau]^{\circ}}}a_k e^{2\pi i(k|x)}\,,
  \qquad x\in\R\,.
\end{gather}
For the two functions $f$ chosen here, it is easy to determine the
corresponding functions $\tilde F$ on the torus $\T = (\R \times \R)/\wt L$
and hence to compute the Fourier-Bohr coefficients $a_k$ exactly by \eqref{coeff},
see \S\ref{exact} . For the
nearest neighbour function these are given explicitly in \eqref{FC}.

On the other hand, the approximation depends on the choice of the refinement
lattice $\wt{L_N}$, which shall
here always be of the form $(1/N)\wt L = (1/N) \wt{\Z [\tau]}$ (so the index
$[\wt{L_N}: \wt L]$ is $N^2$). This determines
both the points that will eventually be projected into our data points and also
the values of $k$ which shall be included in approximating the sum
\eqref{FBexp}, namely a set $\tilde K$ of $\tilde k = (k,k')  \in \wt{\Z[\tau]}^\circ $
chosen as representatives of the group $\wt{\Z[\tau]}^\circ / N\wt{\Z[\tau]}^\circ$ dual to the group
$((1/N) \wt{\Z [\tau]})/\wt L$. Our choice is take the $\tilde k$ with
$|k|$ as small as possible, so $K$ consists of elements $k \in \Z[\tau]$,
one for each congruence class of $\Z[\tau]$ modulo $N\Z[\tau]$,
chosen so that $|k|$ is minimal in its class.
We denote by $f^{exact}(x)$ the finite series, taken from \eqref{FBexp},
approximating $f(x)$, $x\in\R$:
\begin{equation} \label{FBexact}
f^{exact}(x)=\sum_{k\in K} a_k e^{2\pi i(k|x)}\,,
  \qquad x\in\R\,.
\end{equation}

Next we replace the coefficients $a_k$ in $f^{exact}(x)$ by their approximations
\begin{gather}\label{int}
a_k^{int} :=\frac1R\int_0^R f(x)e^{-2\pi i(k|x)}{\rm d}x
\end{gather}
following from~(\ref{coeff2}), and denote the resulting function by
$f^{int}(x)$. We shall use various values of $R$, all of which correspond to
a set of complete passes across the fundamental region, as illustrated in Fig.~5.

The integrals \eqref{int}  are to be estimated by reducing them to finite sums where the
integrand is computed on the finite set of data
arising as projections $\{u_1, \dots, u_{N^2}\}$ of the $N^2$ points in $C$, as explained
in \S\ref{6steps}:
\begin{gather}\label{sum}
a_k^{sum} :=\frac{1}{N^2}\sum_{j=1}^{N^2} f(u_j)e^{-2\pi i(k|u_j)}.
\end{gather}

\smallskip

The resulting approximation to $f^{exact}(x)$ is denoted $f^{sum}(x)$.
This is the approximation that we have been working towards.
Written out in full it reads:
\begin{equation} \label{finalApprox}
f^{sum}(x) = \sum_{k\in K} \,\left\{ \frac{1}{N^2}\sum_{j=1}^{N^2} f(u_j)e^{-2\pi i(k|u_j)}
\right\}\,
e^{2 \pi i (k|x)} \,.
\end{equation}
 It is computed out of data points in $\R$, is a finite sum of exponential functions whose frequencies come
from the Fourier module of $f$, and utilizes discrete groups arising out of the periodic setting
of the underlying cut and project scheme.  By its form, $f^{sum}$ is almost periodic
and approximates the function $\tilde F$ everywhere on the real line as it
lies wrapped around the torus.

Thus we have four functions: $f, \;f^{exact},\; f^{int},\; f^{sum}$, all defined for all $x\in\R$.
The three approximation functions depend on the number $M$ of passes of the real line going
through the fundamental region $C$ (see \S~\ref{ssec_discr_of_analysis}~and~\ref{ssec_restric_to_f} for details),
and on the number $N^2$ of lattice points of $\widetilde{L_N}$ found in $C$. Each of them restricts the full summation
of the Fourier-Bohr expansion to the same finite set $K$ of frequencies.
They differ due to the ways in which the Fourier-Bohr coefficients are obtained:
they are exact in the first case (or at least as exact as real computation on computers can be),
are derived from the integral approximation in the second, and come from the
finite sum approximation to the integral in the third. The calculations and graphs shown below
allow one to compare these functions. The exact coefficients and their
approximations by integrals are sufficiently close as to make little difference to the graphs,
so  the figures are restricted to comparing $f, f^{int}, f^{sum}$.

By way of comparison, there is one further Fourier approximant, of the type that we are
obtained by periodically extending $f$
from its values on a fixed finite interval $I$ of $\R$ to the entire space $\R$.
Thus we introduce
\begin{gather}\label{fcos}
f^{cos}(x)=\sum_{k=0}^n a_k \cos(\frac1{2}{k\pi x}),\qquad
\text{where}\qquad
a_k=\frac 12\int_0^2f(x)\cos(\frac1{2}{k\pi x}){\rm d}x\,.
\end{gather}
Needless to say, $f^{cos}$ cannot be expected to have much relationship to
$f$ outside $I$, and it doesn't. In fact, $f^{cos}$ does not
appear to be a good choice even on the limited domain $I$.




\subsection{Computing the exact Fourier-Bohr coefficients} \label{exact}
It may be of interest to indicate how we computed the exact Fourier-Bohr
coefficients.
We can reorganize Fig.~3, translating
the $a$ and $b$ regions into single blocks (see Fig.~7), a procedure that is familiar
from the klotz construction see $\S$7 often used in studying cut and project sets,
\cite{Kramer1}.

It is then straightforward to understand the corresponding function $\tilde F$
on the 2-dimensional torus, i.e. a genuinely periodic function for which $f(t)=\wt F(t,0)\mod{\wt{\Z[\tau]}}$.

\begin{figure}[h]
\centering
\includegraphics[scale=0.7]{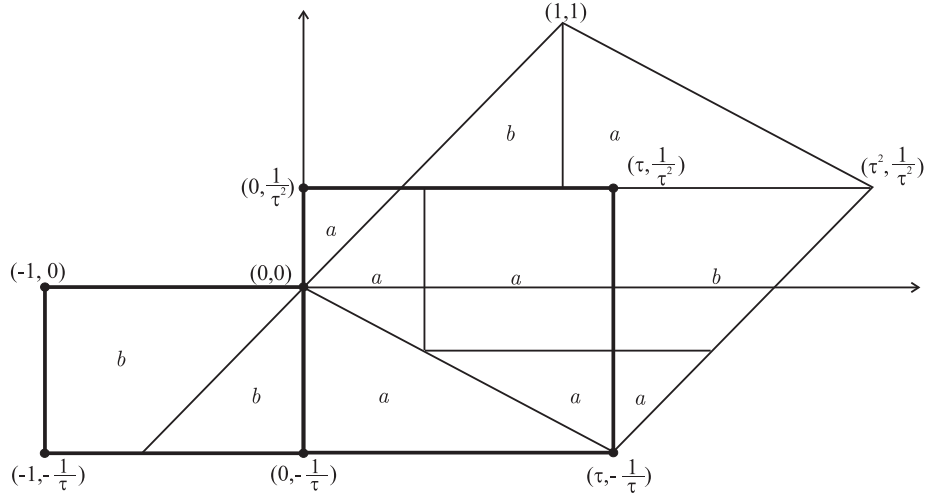}
\caption{ The $a$ and $b$ regions of Fig.~3 are reorganized here into two rectangles (heavy lines).
The function $\wt F$ on $\T$ arising from the local function \eqref{testfunction} is supported on these two rectangles.
It is constant on vertical lines.
Along the horizontal axis, on the $b$ rectangle the function values start at $0$,
increase linearly to $1/2$ at the mid point of the rectangle and then linearly decrease to $0$ again.
Similarly on the $a$ rectangle they increase to $\tau/2$ and then return to $0$. }
\end{figure}

Furthermore, it is easy to compute the `exact' Fourier coefficients of $\widetilde F$.
Let $\wt k=a\wt\omega_1+b\wt\omega_2\in\wt{\Z[\tau]}^\circ$, where $a,b\in\Z$. Then
\begin{gather*}
(\wt k\mid(x,y))=x\,(\wt k\mid(1,0))+y\,(\wt k\mid(0,1))
                  =2xk\delta+2yk'\delta'\,,
\end{gather*}
where
\begin{gather*}
k=\frac{a+b\tau}2\,,\qquad k'=\frac{a+b\tau'}2\,.
\end{gather*}
The Fourier coefficient for $\wt k$ is
\begin{equation}
a_ k = \int_\T \wt F(w) \,e^{-2\pi i(\wt k\mid w)}\,dw
     =\int_0^1\int_0^1\wt F(u,v) \,
           e^{-2\pi i(\wt k\mid u\wt1+v\wt\tau)}\,du\,dv \, .
\end{equation}
 Writing
\begin{gather*}
u\wt1+v\wt\tau=(u+v\tau,u+v\tau')=:(x,y)\,,
\end{gather*}
we have
\begin{gather*}
(\wt k\mid u\wt1+v\wt\tau)=2kx\delta+2k'y\delta'\,.
\end{gather*}
Making the change from the variables $u$ and $v$ to $x$ and $y$, and using the
definition of $\wt F$, we obtain
\begin{eqnarray} \label{FC}
a_k&=&
  \frac1{\sqrt5}  \int_{-\tfrac1\tau}^{0} \int_{-1}^0
  \left(\frac{1}{2} - \left|x + \frac{1}{2}\right| \right)
     e^{-4\pi i(kx\delta+k'y\delta')} \; dx\;dy \nonumber \\
     &+& \frac1{\sqrt5} \int_{-\tfrac1\tau}^{\tfrac1{\tau^2}} \int_{0}^\tau
     \left(\frac{\tau}{2} - \left|x - \frac{\tau}{2}\right| \right)
     e^{-4\pi i(kx\delta+k'y\delta')} \; dx\;dy\,.
\end{eqnarray}

\noindent
Note that the factor $\tfrac1{\sqrt5}$ comes from $x=u+v\tau$, $y=u+v\tau'$.
\begin{gather*}
\left|\begin{matrix} 1&\tau\\1&\tau'\end{matrix}\right|
=\tau'-\tau=-\sqrt5.
\end{gather*}

\subsection{Shadows of singularities}\label{sos}
Inspection of the aperiodic approximations shows that they are remarkably faithful to the originals.
But there is apparently a strange fuzziness
in the approximation in the interval $[-\tau^2 , 0]$. The explanation for this is quite
interesting and shows that the approximation method we use here
is sensitive to quite subtle qualities of the aperiodicity.

The particular Fibonacci set that we have used in our example is {\em singular}, that is to say,
it is one at which the torus parametrization is not one-one. As pointed out in footnote of
\ref{TorusConfusion}, $\Lambda$ is but one of two distinct sets in $X(\Lambda)$
that map to the same point on the torus $\T$ under the torus map.
The other one differs from  $\Lambda$ only in that it contains $\{-\tau\}$ but not $\{-1\}$.
The Fourier analysis takes place on $\T$ and treats these two sets equally.
However the nearest neighbour functions of the two sets are different. This is a difference that it immaterial to
$\tilde F$, which our function $f^{sum}$ is approximating, but which the original
function $F$ can see.  What we see is the Fourier analysis hedging between the two scenarios.

Further refinements of the lattice will never improve the situation
on this interval. However, if we had used an element of $X(\Lambda)$
at which the torus map were $1-1$ (and this is the case with probabilistic certainty if one chooses
randomly from $X(\Lambda)$) then this phenomenon would not occur and
the approximation would be uniformly valid over the entire space.
All one need do is to shift the window so that its end points do not lie in the set $\Z[\tau]$, e.g.
 $\frac{1}{2} + [-1, \frac{1}{\tau}]$.

\bigskip
\section{Final Comments}\label{finalComments}

We should point out that there is considerable scope for adapting the scenario sketched out here.
It is possible to arrange things so that they are related to tilings.
One elegant tiling method is the klotz construction \cite{Kramer1, Kramer2}.
One begins with the
decomposition into Voronoi cells of the lattice $\wt L$ in $\R^d \times \R^d$,
and along with it the corresponding dual cell decomposition into Delone cells.
For each pair of $P,Q$ consisting of intersecting $d$-dimensional faces $P$ and $Q$
from a Voronoi cell and a Delone cell respectively, form the \emph{klotz}
$P^{||} \times Q^\perp$ in $\R^d \times \R^d$. The set of these
kl\"otze form a tiling of $\R^d \times \R^d$.
Furthermore the intersection of $(\R^d,0)$ with this tiling produces a tiling of physical space.
(If one chooses instead to do the projections the other way around, one gets a different tiling).
Several of these kl\"otze can be combined to form a fundamental region for $\R^d \times \R^d$, as we
see in Fig. 7, and this type of choice should lead to computational methods that are adapted to the these tilings.
For instance, the region $A$ of integration might well be chosen as the union of a finite number of tiles.
For more on determining
Voronoi and Delone cells in the context of high symmetry, see \cite{MPvor}.

The method advocated here is based on the idea of local functions,
the extending of them into the context of compact Abelian groups,
and the discretization of the resulting Fourier analysis by the use of finite groups.
In the 1D setting that was explored in detail here, the only symmetries involved arise
from translational symmetry (which is at the base of the almost periodicity).
In higher dimensions, especially those of interest to the quasicrystal community,
decagonal, icosahedral, or other symmetries appear. In these cases there are a number
of ways of utilizing the symmetry to greatly improve the efficiency of the computation,
in the same spirit as \cite{MP}. As we have pointed out, the preparations required for this depend
largely on the cut and project scheme and not so much on the actual model set involved.
Fortunately the cut and project schemes for these settings are essentially canonical
and their strong algebraic nature makes this program quite feasible.

Finally, although we have not spelled it out here, the way in which finite groups and their duals
are used here makes the process amenable to the technique of the fast Fourier transform. Details will appear later.



\subsection*{Acknowledgements}

Work supported in part by the Natural Sciences and Engineering Research Council of Canada,
the MIND Research Institute of Santa Ana, Calif., the Aspen Center for Physics, and MITACS.
The authors are grateful to the
referees for their constructive comments.


\bigskip

R.V.~Moody:
Department of Mathematics, University of Victoria,
Victoria, British Columbia, Canada; rmoody@uvic.ca.

M.~Nesterenko:
         Institute of Mathematics,
         NAS of Ukraine,
         3~Te\-re\-shchen\-kiv\-s'ka Street,
         Kyiv-4, 01601, Ukraine; maryna@imath.kiev.ua.

J.~Patera:
Centre de recherches math\'ematiques,
         Universit\'e de Montr\'eal,
         C.P.6128-Centre ville,
         Montr\'eal, H3C\,3J7, Qu\'ebec, Canada; patera@crm.umontreal.ca.

\newpage
\appendix
\section{Numerical and graphical data for two exact examples of section~\ref{TwoEE},
i.e.,~the~distance-to-the-nearest neighbour function given in~(41) and the~local function defined by~(42)}

\bigskip
\begin{table}[ht]
\caption{Comparison of the calculated coefficients $a_k$ in the approximations
$f^{exact}$, $f^{int}$, and $f^{sum}$ based on the parameters~of~Fig.~10.}
\begin{tabular}{|c|l|l|l|}
\hline&&&\\[-10pt]
$k$ & \multicolumn{1}{|c|}{$a_k^{exact}$} &  \multicolumn{1}{|c|}{$a_k^{int}$}  &  \multicolumn{1}{|c|}{$a_k^{sum}$}\\[2pt]
\hline&&&\\[-10pt]
$-\frac 12-\frac 12\tau$&{\small -0.1065-0.03668i}  &{\small -0.1065-0.0371i}  &{\small -0.1086-0.0581i}\\[2pt]
\hline&&&\\[-10pt]
$-\frac 12$&{\small 0.0243+0.0287i}  &{\small 0.0236+0.0292i}  &{\small 0.0269+0.0711i}\\[2pt]
\hline&&&\\[-10pt]
$-\frac 12+\frac 12\tau$&{\small 0.0026+0.0153i}  &{\small 0.0035+0.0155i}  &{\small 0.0233-0.0332i}\\[2pt]
\hline&&&\\[-10pt]
$-\frac 12\tau$&{\small -0.0683+0.0407i}  &{\small -0.0680+0.0412i}  &{\small -0.0517+0.0542i}\\[2pt]
\hline&&&\\[-10pt]
$0$&{\small 0.3618}  &  {\small 0.3618}  &{\small 0.3367}\\[2pt]
\hline&&&\\[-10pt]
$\frac 12\tau$&{\small -0.0683-0.0407i}  &{\small -0.0680-0.0412i}  &{\small -0.0517-0.0542i}\\[2pt]
\hline&&&\\[-10pt]
$\frac 12-\frac 12\tau$&{\small 0.0026-0.0153i} &{\small 0.0035-0.0155i}  &{\small 0.0233+0.0332i}\\[2pt]
\hline&&&\\[-10pt]
$\frac 12$&{\small 0.0243-0.0287i}  &{\small 0.0236-0.0292i}  &{\small 0.0269-0.0711i}\\[2pt]
\hline&&&\\[-10pt]
$\frac 12+\frac 12\tau$&{\small -0.1065+0.0367i}  &{\small -0.1065+0.0371i}  &{\small -0.1086+0.0581i}\\[2pt]
\hline
\end{tabular}
\end{table}


\begin{table}[ht]
\caption{Comparison of the values of the function $f$ of~\eqref{testfunction} and
its approximants $f^{exact}$, $f^{int}$, $f^{sum}$, and $f^{cos}$, see Fig.~10.}
\begin{tabular}{|c|l|l|l|l|l|}
\hline&&&&\\[-10pt]
$x_i$&$f(x_i)$&$f^{exact}(x_i)$&$f^{int}(x_i)$&$f^{sum}(x_i)$&$f^{cos}(x_i)$\\[2pt]
\hline&&&&&\\[-10pt]
{\small-100}&{\small0.8065}&{\small 0.6916 }&{\small0.6965}&{\small0.7728}&{\small0.1859}\\[2pt]
\hline&&&&&\\[-10pt]
{\small-50}&{\small0.4033}&{\small 0.4229 }&{\small0.4208}&{\small0.4562}&{\small0.3690}\\[2pt]
\hline&&&&&\\[-10pt]
{\small-15}&{\small0.3262}&{\small 0.2555 }&{\small0.2522}&{\small0.1461}&{\small0.4912}\\[2pt]
\hline&&&&&\\[-10pt]
{\small$-3-5\tau$}&{\small0}&{\small 0.0577 }&{\small0.0584}&{\small0.1378}&{\small0.5365}\\[2pt]
\hline&&&&&\\[-10pt]
{\small0}&{\small0}&{\small 0.0658 }&{\small0.0670}&{\small0.1165}&{\small0.1859}\\[2pt]
\hline&&&&&\\[-10pt]
{\small$\tau $}&{\small0}&{\small 0.0797 }&{\small0.0788}&{\small0.0946}&{\small0.1858}\\[2pt]
\hline&&&&&\\[-10pt]
{\small$0.25+\tau$}&{\small0.2500}&{\small 0.2060 }&{\small0.2049}&{\small0.1995}&{\small0.3065}\\[2pt]
\hline&&&&&\\[-10pt]
{\small$0.5+\tau$}&{\small0.5000}&{\small 0.3318 }&{\small0.3313}&{\small0.3416}&{\small0.3138}\\[2pt]
\hline&&&&&\\[-10pt]
{\small$1+\tau$}&{\small0}&{\small 0.1659 }&{\small0.1649}&{\small0.1115}&{\small0.3000}\\[2pt]
\hline&&&&&\\[-10pt]
{\small$1+1.25\tau$}&{\small0.4045}&{\small 0.3325 }&{\small0.3287}&{\small0.1467}&{\small0.5022}\\[2pt]
\hline&&&&&\\[-10pt]
{\small$1+2.5\tau$}&{\small0.8090}&{\small 0.6562 }&{\small0.6609}&{\small0.6949}&{\small0.4681}\\[2pt]
\hline&&&&&\\[-10pt]
{\small$1+2.75\tau$}&{\small0.4045}&{\small 0.3119 }&{\small0.3152}&{\small0.2659}&{\small0.2659}\\[2pt]
\hline&&&&&\\[-10pt]
{\small50}&{\small0.4033}&{\small 0.3265 }&{\small0.3229}&{\small0.1209}&{\small0.3690}\\[2pt]
\hline&&&&&\\[-10pt]
{\small100}&{\small0.1885}&{\small 0.3006 }&{\small0.3004}&{\small0.2282}&{\small0.1859}\\[2pt]
\hline&&&&&\\[-10pt]
{\small500}&{\small0.4396}&{\small 0.4669 }&{\small0.4651}&{\small0.5364}&{\small0.1859}\\[2pt]
\hline
\end{tabular}
\end{table}

\begin{figure}[t]
\centering
\includegraphics[scale=0.7]{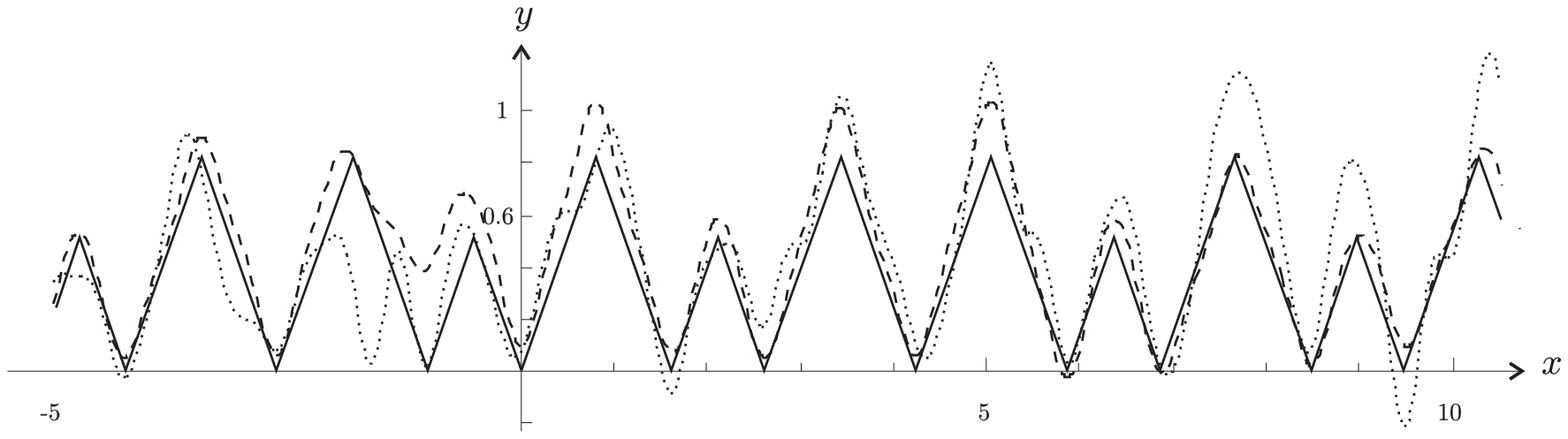}
\caption{The solid line is the graph of $f$ of \eqref{testfunction}.
The dotted curve shows the $f^{sum}$ approximation
and the dashed curve shows the $f^{int}$ approximation.
Both approximations were calculated using a total of $N^2=49$ lattice points in $C$
and $M=11$ path passes in $C$,
corresponding to an interval of integration  $R\approx23,30$.}

\vspace{18pt}

\centering
\includegraphics[scale=0.7]{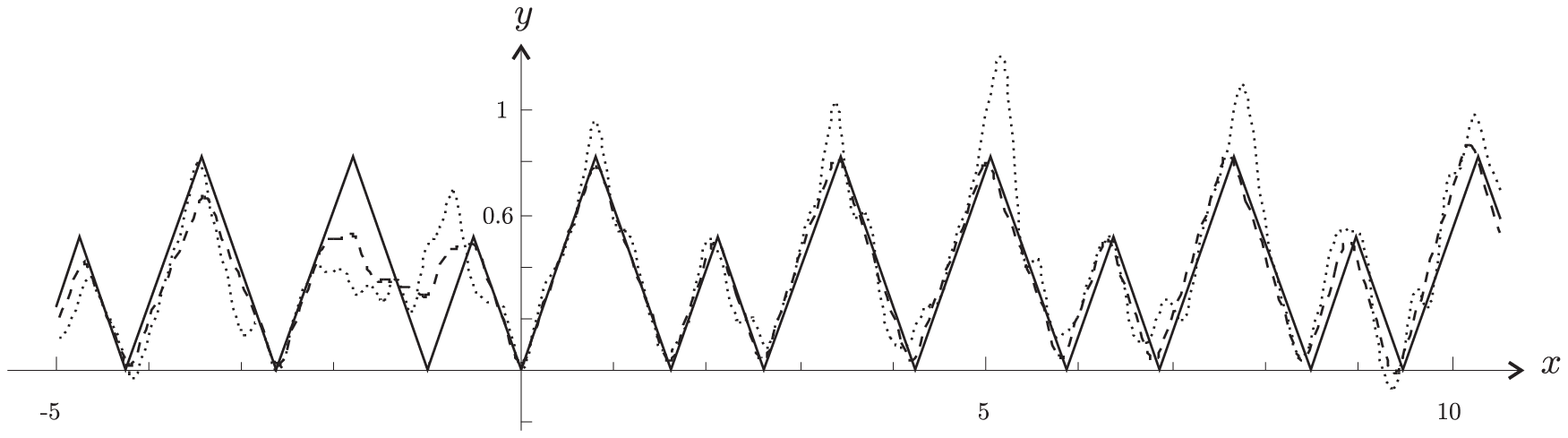}
\caption{The function $f$ and approximants of \eqref{testfunction}
are drawn with the same conventions as in  Fig.~8, but for refined
parameters:  $M=17$, $N^2=121$, and $R\approx37,43$. See
subsection~\ref{sos} for an explanation of the misfit of
approximations in the region $[-\tau^2,0]$.}

\vspace{18pt}

\centering
\includegraphics[scale=0.7]{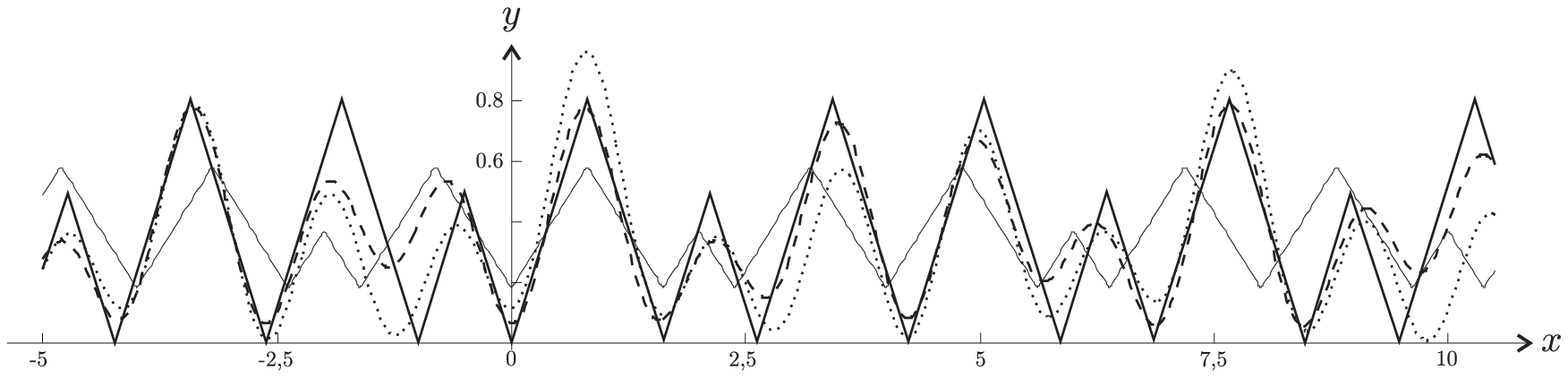}
\caption{The function $f$ of \eqref{testfunction} is graphed for parameters:
$N^2=9$ lattice points from $C$, $M=10$ passes of the fundamental domain, and corresponding
interval of integration given by
$R\approx21,64$. Also the periodic approximation $f^{cos}$ is
shown. The number of cosine terms used is $n=50$ and coefficients are calculated by formula \eqref{fcos}.
Here the dotted curve corresponds to the $f^{sum}$ approximation, the dashed curve
corresponds to $f^{exact}$ approximation, and the thin line represents $f^{cos}$ approximation.
}
\end{figure}


\begin{figure}
\centering
\includegraphics[scale=0.7]{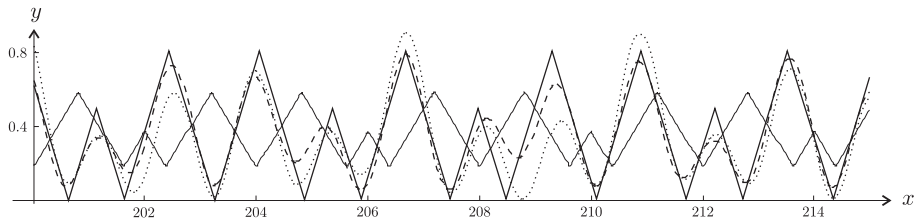}
\caption{Here we graph the same approximations as in Fig.~10, but
the functions are now drawn on the interval $[200, 215]$. Observe
how the aperiodic approximants continue to follow the graph of $f$ while
the periodic approximant now has no relation to it.}
\end{figure}

\begin{figure}
\centering
\includegraphics[scale=0.7]{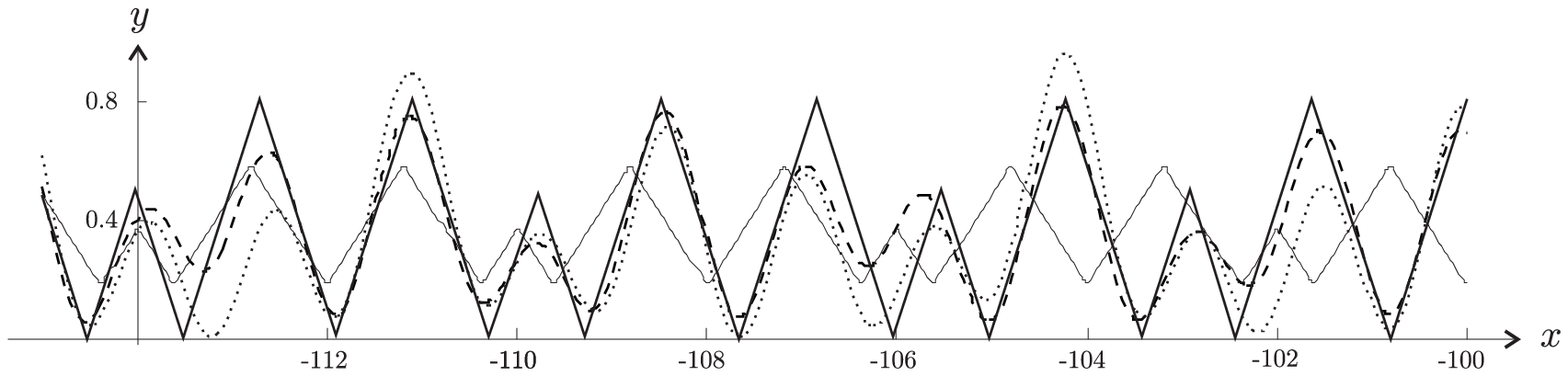}
\caption{The approximations are presented with the same conventions as on Fig.~10
but the functions are drawn in the interval $[-115, -100]$.}
\end{figure}


\begin{table}
\caption{Comparison of the calculated coefficients $a_k$ in the approximations
$f^{exact}$, $f^{int}$, and $f^{sum}$. The parameters are as in Fig.~14.}
\begin{tabular}{|c|l|l|l|}
\hline&&&\\[-10pt]
$k$ & \multicolumn{1}{|c|}{$a_k^{exact}$} &  \multicolumn{1}{|c|}{$a_k^{int}$}  &  \multicolumn{1}{|c|}{$a_k^{sum}$}\\[2pt]
\hline&&&\\[-10pt]
$-\frac 12-\frac 12\tau$&{\small-0.1065-0.0367i}  &{\small-0.1065-0.0371i}  &{\small-0.1086-0.0581i}\\[2pt]
\hline&&&\\[-10pt]
$-\frac 12$&{\small 0.0243+0.0287i}  &{\small0.0236+0.0292i}  &{\small0.0269+0.0711i}\\[2pt]
\hline&&&\\[-10pt]
$-\frac 12+\frac 12\tau$&{\small 0.0026+0.0153i}  &{\small0.0035+0.0155i}  &{\small0.0233-0.0332i}\\[2pt]
\hline&&&\\[-10pt]
$-\frac 12\tau$&{\small-0.0683+0.0407i}  &{\small-0.0680+0.0412i}  &{\small-0.0517+0.0542i}\\[2pt]
\hline&&&\\[-10pt]
$0$&{\small0.3618}  &{\small0.3618}  &{\small0.3367}\\[2pt]
\hline&&&\\[-10pt]
$\frac 12\tau$&{\small-0.0683-0.0407i}  &{\small-0.0680-0.0412i}  &{\small-0.0517-0.0542i}\\[2pt]
\hline&&&\\[-10pt]
$\frac 12-\frac 12\tau$&{\small0.0026-0.0153i}  &{\small0.0035-0.0155i}  &{\small0.0233+0.0332i}\\[2pt]
\hline&&&\\[-10pt]
$\frac 12$&{\small0.0243-0.0287i}  &{\small0.0236-0.0292i}  &{\small0.0269-0.0711i}\\[2pt]
\hline&&&\\[-10pt]
$\frac 12+\frac 12\tau$&{\small-0.1065+0.0367i}  &{\small-0.1065+0.0371i}  &{\small-0.1086+0.0581i}\\[2pt]
\hline
\end{tabular}
\end{table}

\begin{table}
\caption{Comparison of the values of the function $f$ of~\eqref{testfunction2} and
its approximants $f^{exact}$, $f^{int}$, $f^{sum}$, and $f^{cos}$.}
\begin{tabular}{|c|l|l|l|l|l|}
\hline&&&&&\\[-10pt]
$x_i$&$f(x_i)$&$f^{exact}(x_i)$&$f^{int}(x_i)$&$f^{sum}(x_i)$&$f^{cos}(x_i)$\\[2pt]
\hline&&&&&\\[-10pt]
{\small-100}&{\small 1}&{\small 1.0960 }&{\small1.0796}&{\small1.1907}&{\small0.8024}\\[2pt]
\hline&&&&&\\[-10pt]
{\small-50}&{\small 1}&{\small 1.0690 }&{\small1.0628}&{\small0.6269}&{\small-0.1948}\\[2pt]
\hline&&&&&\\[-10pt]
{\small-15}&{\small 1}&{\small 1.1020}&{\small1.1637}&{\small1.1570}&{\small0.8162}\\[2pt]
\hline&&&&&\\[-10pt]
{\small$-3-5\tau$}&{\small 1}&{\small 0.1092 }&{\small0.0963}&{\small0.3515}&{\small0.8093}\\[2pt]
\hline&&&&&\\[-10pt]
{\small0}&{\small 1}&{\small 0.5331 }&{\small0.5268}&{\small1.5952}&{\small0.8024}\\[2pt]
\hline&&&&&\\[-10pt]
{\small$\tau $}&{\small -1}&{\small -0.0271 }&{\small-0.0291}&{\small0.7757}&{\small0.3036}\\[2pt]
\hline&&&&&\\[-10pt]
{\small$0.25+\tau$}&{\small -1}&{\small -1.2750 }&{\small-1.2957}&{\small-1.3115}&{\small-0.1820}\\[2pt]
\hline&&&&&\\[-10pt]
{\small$0.5+\tau$}&{\small -1}&{\small -0.7919 }&{\small-0.8123}&{\small-0.7404}&{\small-0.1884}\\[2pt]
\hline&&&&&\\[-10pt]
{\small$1+\tau$}&{\small 1}&{\small 0.0704 }&{\small0.0971}&{\small-0.0128}&{\small0.7910}\\[2pt]
\hline&&&&&\\[-10pt]
{\small$1+1.25\tau$}&{\small 1}&{\small 0.9261 }&{\small0.9625}&{\small1.1215}&{\small0.8005}\\[2pt]
\hline&&&&&\\[-10pt]
{\small$1+2.5\tau$}&{\small 1}&{\small 1.1233 }&{\small1.1382}&{\small1.0246}&{\small 0.7991}\\[2pt]
\hline&&&&&\\[-10pt]
{\small$1+2.75\tau$}&{\small 1}&{\small 0.9577 }&{\small0.9519}&{\small1.2470}&{\small 0.7826}\\[2pt]
\hline&&&&&\\[-10pt]
{\small50}&{\small 1}&{\small 0.9241 }&{\small0.9374}&{\small1.1735}&{\small-0.1948}\\[2pt]
\hline&&&&&\\[-10pt]
{\small100}&{\small -1}&{\small -1.2440 }&{\small-1.2652}&{\small-1.5375}&{\small 0.8024}\\[2pt]
\hline&&&&&\\[-10pt]
{\small500}&{\small 1}&{\small 1.0950 }&{\small 1.0639}&{\small 0.9970}&{\small 0.8024}\\[2pt]
\hline
\end{tabular}
\end{table}

\begin{figure}[h]
\centering
\includegraphics[scale=0.7]{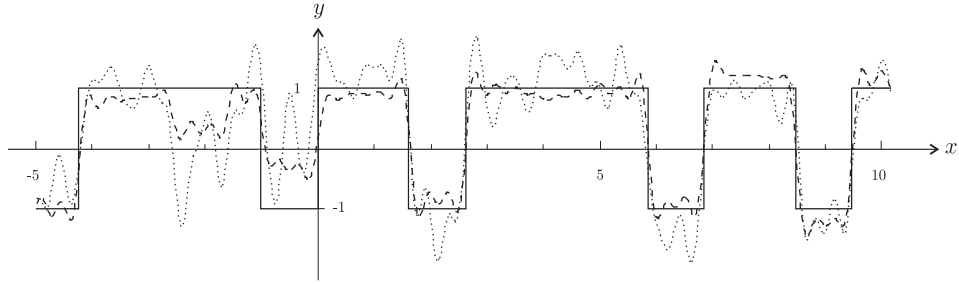}
\caption{
The solid line is the graph of $f$ of \eqref{testfunction2}.
The dotted and dashed curves correspond to the $f^{sum}$ and $f^{int}$ approximations.
Both approximations were calculated using a total of $N^2=81$ lattice points in $C$, $M=17$ passes of $C$,
and corresponding interval of integration given by
 $R\approx37,01$.
}

\end{figure}

\begin{figure}[h]
\centering
\includegraphics[scale=0.7]{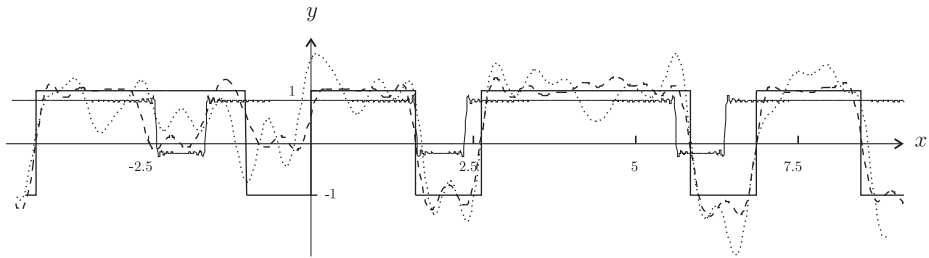}
\caption{
The function $f$ of \eqref{testfunction2} is drawn for parameters:  $N^2 =49$, $M=11$,
$R\approx23,30$. Also the periodic approximation $f^{cos}$ is shown as thin line.
The number of cosine terms used is  $n=50$ and coefficients were calculated by
\eqref{fcos}.
Here the dotted curve corresponds to the $f^{sum}$ approximation and the dashed curve
corresponds to $f^{exact}$ approximation.
}
\end{figure}

\begin{figure}[h]
\centering
\includegraphics[scale=0.7]{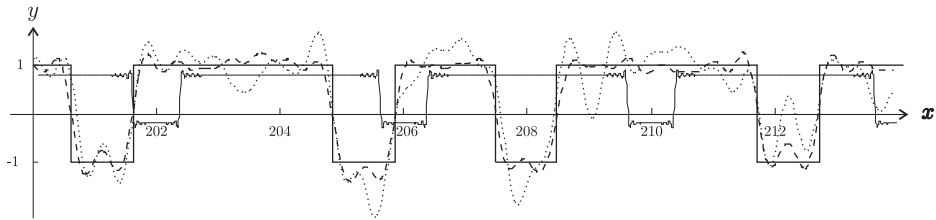}
\caption{Here we adduce the same approximations as in Fig.~14, but
the functions are drawn on the interval $[200, 215]$.}
\end{figure}


\begin{figure}[h]
\centering
\includegraphics[scale=0.69]{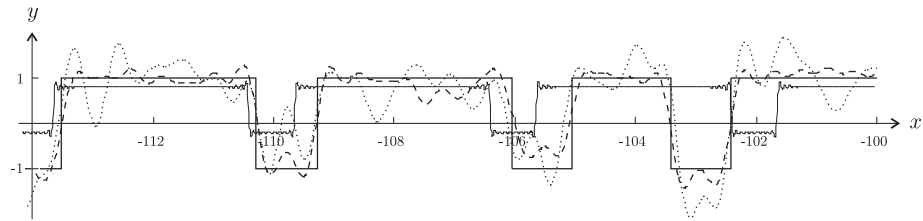}
\caption{The approximations are drawn with the same conventions as on Fig.~14, but
on the interval $[-115,-100]$.}
\end{figure}

\end{document}